\documentclass[preprint,nofootinbib,aps,superscriptaddress,eqsecnum]{revtex4-2} 
\usepackage{float}
\pdfoutput=1
\usepackage{extarrows}
\usepackage{tabularx}
\usepackage{graphicx}
\usepackage{amsmath}
\usepackage{amsfonts}
\usepackage{amssymb}
\usepackage{hyperref}
\usepackage{nameref}
\usepackage[justification=centering]{caption} 
\usepackage[margin=1.5cm]{geometry}
\usepackage{graphicx,subcaption,lipsum}
\let\revappendix\appendix
\usepackage{caption}
\usepackage{subcaption}
\usepackage{color}
\allowdisplaybreaks
\def\bea{\begin{eqnarray}}
	\def\eea{\end{eqnarray}}
\def\be{\begin{equation}}
	\def\ee{\end{equation}}

\begin{document}
	\title{Study of Neutrino Phenomenology and $0\nu\beta\beta$ Decay using Polyharmonic $Maa\beta$ Forms}
	\author{Bhabana Kumar}
	\email{bhabana12@tezu.ernet.in}
	\author{Mrinal Kumar Das}
	\email{mkdas@tezu.ernet.in}
	\affiliation{Department of Physics, Tezpur University, Tezpur 784028, India}

\begin{abstract}
In this study, we explore the application of the $\Gamma_{3}$ modular group, which is isomorphic to the $A_{4}$ symmetric group  in developing a model for neutrino mass. We realized a non-supersymmetric left-right asymmetric model incorporating modular symmetry, where the modular forms consist of both holomorphic and non-holomorphic components and the Yukawa couplings expressed through polyharmonic $Maa\beta$ forms. To effectively implement the extended see-saw process in this model, we introduce one fermion singlet for each generation. We compute the effective mass and the associated half-life of $0\nu\beta\beta$ by accounting for both standard and non-standard contributions.  Additionally, our study investigates the non-unitary effects and CP-violation arising from non-unitarity in this context. The model predicts values for the sum of neutrino masses and neutrino oscillation parameters are consistent with experiments. Furthermore, it yields satisfactory results in calculating the effective mass and half-life of $0\nu\beta\beta$ decay. These findings highlight the possibility and benefit of employing modular symmetry in neutrino mass model construction.
\end{abstract}
\maketitle
\newpage
\section{\textbf{Introduction}}
One of the most recognized and thoroughly tested particle physics models to date is the Standard Model (SM). In addition to explaining interactions between all of the known fundamental particles, SM also describes their properties, such as mass, charge, spin, and so forth. Despite being successful, SM is not without limitations. One significant drawback is its inability to account for the tiny neutrino masses. Except for the neutrino, every fermion in the context of SM is massive, and the Higgs mechanism gives those fermions their mass. The Higgs mechanism requires both the left and right-handed counter particles to couple with the Higgs field for a particle to gain mass. However, due to the lack of right-handed counter particles of neutrinos, it is not possible to incorporate massive neutrinos within the framework of SM. The idea of a massless neutrino is completely denied by the discovery of neutrino oscillation. Experiments like Superkamiokande \cite{Super-Kamiokande:2001bfk} and SNO \cite{SNO:2002hgz,SNO:2002tuh,Bandyopadhyay:2001aa} have provided strong evidence for neutrino oscillation, a phenomenon that is not possible if the neutrino is massless. To learn more about neutrinos, several experiments, including MINOS \cite{Evans:2013pka}, Daya-Bay \cite{DayaBay:2012fng}, TK2 \cite{T2K:2011ypd}, RENO\cite{Lasserre:2012ax}, and others, have been carried out, and all those experiments have suggested non-zero and non-degenerate neutrino masses. Neutrino oscillation experiments have played a crucial role in refining our knowledge of neutrinos and providing insights into their properties. Scientists have been able to determine the mass squared differences and mixing angles associated with neutrinos. While substantial progress has been made in determining the oscillation parameters, there is still much we do not know definitively about neutrinos. The Planck experiment recently provided an approximate upper bound on the sum of the neutrino masses, which is equal to $\sum m_{\nu}=0.11$ eV \cite{Planck:2018vyg}. However, the exact masses of neutrino and their mass hierarchy \cite{Qian:2015waa, Ghosh:2012px} remain unclear. Additionally, the nature of neutrinos, whether they are Dirac or Majorana particles \cite{Barenboim:2002hx, Czakon:1999cd}, is still a subject of ongoing study.\\
The limitations of the SM extend beyond neutrino masses. It fails to explain origin of CP violation in weak interactions, lepton flavor violation, the baryon asymmetry of the universe, dark matter, and dark energy. 
To address these shortcomings, we need to go beyond the SM and explore extensions that involve the addition of new fermion or scalar particles. The seesaw mechanism is a prominent extension that offers explanations to the tiny neutrino mass problem. There are different types of seesaw mechanisms, such as Type I \cite{Mohapatra:1979ia,Gell-Mann:1979vob, Schechter:1980gr}, Type II \cite{Cheng:1980qt,Wetterich:1981bx,Magg:1980ut}, and Type III \cite{Foot:1988aq}. In these models, the SM is extended by incorporating right-handed neutrinos,  $SU(2)_{L}$ scalar triplet, or $SU(2)_{L}$ triplet fermion, respectively, to generate small but nonzero neutrino masses.\\
Another beyond SM  framework is the Left-Right symmetric model (LRSM) \cite{Pati:1974yy,Senjanovic:1975rk,Deshpande:1990ip,Mohapatra:1974gc}, which extends the SM by introducing additional gauge symmetries. The gauge group of the LRSM is $SU(3)_{C} \times SU(2)_{L} \times SU(2)_{R} \times U(1)_{B-L} \times D$($G_{221D}$). This model has gained significant attention in the literature due to its success in explaining the tiny neutrino masses observed experimentally as well as addressing other drawbacks of the SM. Even though the model is very successful in many aspects, ambiguity arises when one wants to incorporate the LRSM within the Grand Unified Theory (GUT). To embed LRSM within GUT like $SO(10)$, requires the parity breaking scale to be very high, only then one will get the observed value of $\sin^{2}\theta_{W}$ \cite{Georgi:1974sy}, but in LRSM both the D-parity and $SU(2)_{R}$ gauge groups break at the same energy scale, which implies that all the phenomena involving right-handed current will get highly suppressed under this condition.  
However, an alternative approach to the LRSM is the left-right asymmetric model \cite{Rizzo:1981dm, Chang:1983fu}, where the D-parity is decoupled from the $SU(2)_{R}$ gauge group at a very high energy scale. This decoupling results in different gauge coupling values for the $SU(2)_{L}$ and $SU(2)_{R}$ gauge groups, leading to an asymmetry in the model. In this left-right asymmetric model, the gauge coupling value of $SU(2)_{L}$ is not equal to the gauge coupling value of $SU(2)_{R}$, that is, $g_{l} \neq g_{r}$. This alternative approach is intriguing because the different gauge coupling values can either suppress or enhance various low-energy phenomena, offering unique predictions.\\
For many years discrete symmetry groups like $A_{4}$, $S_{4}$, $S_{3}$ have been used for model building purposes in the lepton sector \cite{King:2013eh,Altarelli:2010gt,Ma:2001dn,Altarelli:2009gn,Ishimori:2010au,Kobayashi:2018vbk,Chauhan:2023faf,King:2017guk} . But one of the main disadvantages is the requirement of "flavon" to spontaneously break such symmetry and the vacuum expectation value (V.E.V) of such field can significantly affect the physical prediction of the model. In recent years, the application of modular symmetry in particle physics has received significant attention due to its ability to explain the patterns and hierarchies observed in fermions, also modular symmetry gained its popularity because one can construct a model without using any "flavon" since in modular symmetry modulus $\tau$ is responsible for breaking the flavor symmetry \cite{Novichkov:2019sqv,King:2020qaj,deAdelhartToorop:2011re,Feruglio:2017spp}. Many research groups have been working on model building by using modular group like $\Gamma_{3}$ \cite{Behera:2020sfe,Nomura:2019xsb,Gogoi:2022jwf,Kumar:2023moh,Kashav:2022kpk,Kashav:2021zir,Behera:2020lpd,deAnda:2018ecu},$\Gamma_{4}$ \cite{Okada:2019lzv,Novichkov:2020eep,Penedo:2018nmg,Ding:2021zbg}, $\Gamma_{5}$ \cite{Novichkov:2018nkm} etc.\\
In our current work, we have focused on studying neutrino masses and mixing within the left-right asymmetric model. So far, model building using modular symmetry has been done in the framework of supersymmetry due to the holomorphic nature of modular forms\cite{Feruglio:2017spp}. However, in our case, we are working in a non-supersymmetric framework inspired by automorphic forms\cite{Qu:2024rns,Ding:2020zxw}. Here, the concept of holomorphicity is replaced by the Laplacian condition, where holomorphicity is substituted with the harmonic condition and the Yukawa couplings are polyharmonic $Maa\beta$ forms of level N, which can be arranged into multiplets of the finite modular groups $\Gamma_{N}$ and $\Gamma^{\prime}_{N}$.   The level N polyharmonic $Maa\beta$ forms match the level N holomorphic modular forms at weights $k\geq3$. However, this framework allows for the existence of polyharmonic $Maa\beta$ forms with negative and zero weights. In the case of $\Gamma_{3}$ modular group, which is isomorphic to $A_{4}$ group, for $k\leq 2$, we always have four polyharmonic $Maa\beta$ forms among which one is a singlet and the other three forming a triplet of $A_{4}$. So in the present work, we have used the idea of automorphic forms, and we have studied neutrinoless double beta ($0\nu\beta\beta$) decay within this framework. By utilizing this mathematical framework, we have derived mass matrices that govern the neutrino sector of the model and implemented those matrices to calculate and study neutrino masses, effective neutrino mass, and half-life of $0\nu\beta\beta$ decay.\\
The overall structure of the paper is as follows: In Section ~\ref{s2}, we provide a brief introduction to the left-right asymmetric model. In Section~ \ref{s3}, we present an overview of modular symmetry. In Section \ref{s4}, we thoroughly discuss the model. In Section ~\ref{s5}, we provide an overview of the extended inverse seesaw mechanism, while in Section ~\ref{s6}, we discuss the unitary violation in the lepton sector. In Section~ \ref{s7}, we examine the contribution to $0\nu\beta\beta$ decay. A detailed discussion of the numerical analysis and results of our study is presented in Section ~\ref{s8}. Finally, we summarize our findings in Section ~\ref{s9}.
\section{\textbf{Left-right asymmetric model}} \label{s2}
The initial proposal of the left-right asymmetry model was put forth by Chang, Mohapatra, and Parida in their publication \cite{Chang:1984uy}. An alternate method has been proposed to decouple D-parity from the $SU(2)_{R}$ gauge group at high energy scales while preserving the original gauge symmetry \cite{Chang:1983fu,Rizzo:1981dm}. The decoupling of D-parity from the $SU(2)_{R}$ gauge group occurs when the odd parity singlet scalar field $\eta$, acquires a vacuum expectation value (V.E.V) at an energy scale denoted as $M_{P}$. The outcome is an asymmetrical left-right model in which the gauge couplings of $SU(2)_{L}$ and $SU(2)_{R}$ becomes unequal, i.e., $g_{l}\neq g_{r}$. The entire symmetry-breaking steps can be illustrated as follows :
\begin{center} 
	\centering
	$ SU(2)_{L} \times SU(2)_{R} \times U(1)_{B-L} \times  SU(3)_{C} \times D $ \\ $ \downarrow $ \hspace{0.5cm} $\eta$ \\ 
	
	$ SU(2)_{L} \times SU(2)_{R} \times U(1)_{B-L} \times SU(3)_{C}$ \\      
	$\downarrow $ \vspace{0.5cm} $\Sigma$\\ 
	
	$ SU(2)_{L} \times U(1)_{R} \times U(1)_{B-L} \times SU(3)_{C}$ \\ $ \downarrow$ \vspace{0.5 cm} $\Delta_{R}$\\ 
	
	$ SU(2)_{L} \times U(1)_{Y} \times SU(3)_{C}  $ \\
	$\downarrow $\hspace{0.5cm} $\Phi$ \\
	
	$ U(1)_{em} \times SU(3)_{C}$. \\  
	
\end{center}
After the spontaneous breaking of D-parity, the asymmetric gauge group undergoes two possible paths. The first path occurs when the Higgs triplet $\Delta _{R}$ gains a non-zero vacuum expectation value, leading to the breakdown of the left-right asymmetric model into SM gauge group. Subsequently, the SM gauge group further breaks down to $U(1)_{em}$ when both the left-handed Higgs triplet and the bidoublet acquire non-zero V.E.V. \\
On the other hand, in the alternative path, the gauge group $SU(2)_{R}$ initially breaks down to $U(1)_{R}$ when a heavier scalar triplet, $\Sigma (1,3,0)$, carrying $B-L=0$, obtains a non-zero V.E.V. This step of symmetry breaking leads to the generation of massive $W^{\pm}_{R}$ gauge bosons. Finally, in the last step, the intermediate gauge group $U(1)_{R} \times U(1)_{B - L}$ breaks down to SM gauge group when either a doublet, a triplet, or both the doublet and triplet gain non-zero V.E.V \cite{Sruthilaya:2017vui}. During this step, the right-handed neutral gauge boson $Z^{0}_{R}$ obtains its mass. The minimal particle content of the model and their charge assignments under the gauge group $G_{221}$ are given below \\
\begin{align*}
	Q_{L} = \begin{pmatrix}
		u_{L} \\
		d_{L} 
	\end{pmatrix},
	&  \hspace{1cm}
	Q_{R} = \begin{pmatrix}
		u_{R}\\
		d_{R}
	\end{pmatrix}
	& \hspace{1cm}
	\Psi_{L} = \begin{pmatrix}
		\nu_{L} \\
		l_{L} 
	\end{pmatrix},
	&  \hspace{1cm}
	\Psi_{R} = \begin{pmatrix}
		\nu_{R}\\
		l_{R}
	\end{pmatrix}
\end{align*}

$Q_{L}$ : $(2, 1,1/3)$ \hspace{1cm} $Q_{R}$ : $(1,2, 1/3)$ \hspace{1cm}$\Psi_{L}$ : $(2,1,-1)$ \hspace{1cm} $\Psi_{R}$ : $(1,2,-1)$ 

and the required Higgs multiplets are\\
\begin{align*}
	\phi = \begin{pmatrix}
		\phi_{1}^{0} & \phi^{+}_{1} \\
		\phi_{2}^{-} & \phi_{2}^{0}
	\end{pmatrix},
	&  \hspace{1cm}
	\Delta_{L,R} = \begin{pmatrix}
		\frac{\delta_{L,R}^{+}}{\sqrt{2}} & \delta_{L,R}^{++} \\
		\delta_{L,R}^{0} & -\frac{\delta_{L,R}^{+}}{\sqrt{2}}
	\end{pmatrix}
\end{align*}

$\eta (1,1,0)$, \hspace{0.5cm}   $\Phi (2,2,0)$, \hspace{0.5 cm}$\Delta_{L} (3,1,+2) + \Delta_{R}(1,3,+2)$ . \\
The transformation properties of those scalar field under parity are given below
$$\Delta_{L}\rightarrow \Delta_{R} ;~~~ \Phi \rightarrow \Phi^{\dagger};~~~ \eta \rightarrow -\eta$$
We can write the individual potential term as follows \cite{Chang:1984uy}
\begin{gather*}
	\begin{aligned}
		V_{\Delta}&= \mu^{2}_{\Delta}[Tr(\Delta^{\dagger}_{L} \Delta_{L}) + Tr(\Delta^{\dagger}_{R}\Delta_{R})] \\
		V_{\eta}&=-\mu^{2}_{\eta}\eta^{2} + \lambda\eta^{4}  \\
		V_{\eta \Delta}& = M\eta[Tr(\Delta^{\dagger}_{L}\Delta_{L})-Tr(\Delta^{\dagger}_{R}\Delta_{R})] + \lambda_{1}\eta^{2}[Tr(\Delta^{\dagger}_{L}\Delta_{L})+Tr(\Delta^{\dagger}_{R}\Delta_{R})]\\
		V_{\eta \Phi}& =\lambda_{2}\eta^{2} Tr(\phi^{\dagger}\phi) + \lambda_{3}\eta^{2}(det\Phi +det \Phi) \\  
		V_{\Delta \Phi}& = \lambda_{5}[Tr(\Tilde{\Phi)}\Delta_{R}\Phi^{\dagger}\Delta^{\dagger}_{L} + Tr(\Tilde{\Phi^{\dagger}}\Delta_{L}\Phi\Delta^{\dagger}_{R})]\\  
		V_{\Phi}&= \mu^{2}_{1} Tr(\Phi^{\dagger}\Phi) + \mu^{2}_{2}[Tr(\Tilde{\Phi^{\dagger}}\Phi)+Tr(\Tilde{\Phi^{\dagger}}\Phi)]
	\end{aligned}    
\end{gather*}	

After the spontaneous symmetry breaking we can choose the V.E.V of $\Delta_{L,R}$ and $\Phi$ as follows\\
\begin{align*}
	<\Delta_{L,R} > = \frac{1}{\sqrt{2}} \begin{pmatrix}
		0 & 0\\
		\nu_{L,R} & 0
	\end{pmatrix} ,
	&  ~~
	<\Phi> = \begin{pmatrix}
		k & 0\\
		0 & \exp{(i\phi)}k^{\prime}
	\end{pmatrix}
\end{align*}
and the D-parity is broken by $< \eta >=\frac{\mu}{\sqrt{2\lambda}}$, which then makes the $\Delta_{L}$ and $\Delta_{R}$ mass terms asymmetric. 
\section{\textbf{Modular Group}} \label{s3}
The modular group $SL(2,Z)=\Gamma$ is defined as a group of $2\times2$ matrices with positive or negative integer element and determinant equal to $1$ and it represents the symmetry of a torus \cite{deAnda:2023udh,deAdelhartToorop:2011re}. It is infinite group and generated by two generators of the group $S$ and $T$.\\
\begin{equation}\label{Q1}
	\Gamma = \Biggl\{\begin{pmatrix}
		a & b\\
		c & d
	\end{pmatrix} | a,b,c,d \in \mathbb{Z}, ad-bc=1\Biggr\}~~ .
\end{equation}
The generator of the group satisfy the conditions:\\
$S^{2} = 1$ \hspace{0.5cm} and  \hspace{0.5cm} $(ST)^{3} = 1$ \\
and they can be represented by $2\times 2$ matrices 
\begin{align*}
	S=\begin{pmatrix} 
		0 & 1\\
		-1 & 0 \\
	\end{pmatrix},
	&  ~~~~~~
	T= \begin{pmatrix}
		1 & 1 \\
		0 & 1 \\
	\end{pmatrix} ~~~.
\end{align*}
A two-dimensional space is obtained, when the torus is cut open and this two-dimensional space can be viewed as an Argand plane and modulus $\tau$ is the lattice vector of that Argand plane. The transformation of modulus $\tau$ \cite{Ferrara:1989qb} of the modular group on the upper half of the complex plane is given below 
\begin{equation*}
	\gamma : \tau \rightarrow \gamma(\tau) = \frac{(a\tau + b)} {(c\tau + d)}
\end{equation*} 
the transformation of $\tau$ is same for both $\gamma$ and $-\gamma$ and we can define a group $\bar{\Gamma}= PSL(2,Z)$, which is a  projective special linear group. Also, the modular group has an infinite number of normal subgroups, which is the principal congruence subgroup of level N and can be defined as
\begin{equation}\label{Q2}
	\Gamma(N)= \Biggl\{\begin{pmatrix}
		a & b\\
		c & d
	\end{pmatrix} \in SL(2,Z), \begin{pmatrix}
		a & b \\
		c & d
	\end{pmatrix}= \begin{pmatrix}
		1 & 0 \\
		0 & 1
	\end{pmatrix} (mod N)\Biggr\},
\end{equation}  
and for $N>2$, $\bar{\Gamma}(N)=\Gamma_{N}$. The use of a finite group is essential for the purposes of model building. Usually a modular group is an infinite group but we can obtain a finite modular group for $N>2$, if we consider the quotient group $\Gamma_{N} =PSL(2,Z)/\bar{\Gamma}(N)$ and these modular group are isomorphic to non-abelian discrete groups. The modular invariance requires the Yukawa couplings to be a certain modular function $Y(\tau)$ and should follow the following transformation property.
\begin{equation}\label{n2}
		Y(\gamma\tau)=(c\tau+d)^{k}Y(\tau)
\end{equation} 
So far modular symmetry has been used in the context of supersymmetry where the superpotential is the holomorphic function of modulus $\tau$. However, some recent works have been developed where we can use modular symmetry to develop the non-supersymmetric model. The basic idea is that one can realise the non-supersymmetric framework by using the framework of automorphic forms and the assumption of holomorphicity is replaced by the Laplacian condition. In such case the Yukawa coupling can have both holomorphic and non-holomorphic parts \cite{Qu:2024rns,Ding:2020zxw,Ding:2024inn}. In the present work, we are concerned with the polyharmonic $Maa\beta$ forms of weight k and the Yukawa coupling needs to follow another transformation property, which is given below
\begin{equation}\label{n3}
		\Delta_{k}Y(\tau)=0
\end{equation}
where $\tau=x+iy$ and $\Delta_{k}$ is the hyperbolic Laplacian operator
	\begin{equation}
		\Delta_{k}=-y^{2}\big(\frac{\partial^{2}}{\partial x^{2}} + \frac{\partial^{2}}{\partial y^{2}}\big) + iky\big(\frac{\partial}{\partial x}+ i\frac{\partial}{\partial y}\big) = -4y^{2} \frac{\partial}{\partial \tau}\frac{\partial}{\partial \tau}+ 2iky \frac{\partial}{\partial \tau}
\end{equation}
The weight k of polyharmonic $Maa\beta$ forms can be positive, zero, or even negative. Based on the transformation property given in the equation \eqref{n2}, which implies that $Y(\tau + N)=Y(\tau)$ and considering the transformation property from equation \eqref{n3}, the Fourier expansion of a level N and weight k polyharmonic $Maa\beta$ form can be expressed as\cite{Qu:2024rns}
\begin{equation}\label{n01}
		Y(\tau)= \sum_{n\in\frac{1}{N}\mathbb{Z} n\geq 0} c^{+}(n)q^{n} + c^{-}(0)y^{1-k} + \sum_{n\in \frac{1}{N}\mathbb{Z}n< 0} c^{-}(n)\Gamma(1-k,-4\pi ny)q^{n}
\end{equation}
where $q=e^{i2\pi\tau}$. \\
Table \ref{w1t1} shows the summary of polyharmonic $Maa\beta$ forms of weights $ k_{Y} = -4, -2, 0, 2, 4, 6$ at level  $N = 3$\cite{Qu:2024rns}.\\
\begin{table}[ht]
	\centering
	\begin{tabular}{|m{4cm}|m{6cm}|}
		\hline
		\textbf{Weight $k_{Y}$} & \textbf{Polyharmonic $Maa\beta$ forms $Y^{k_{Y}}_{r}$} \\ \hline
		$k_{Y}=-4$ & $Y^{-4}_{1}$,~~$Y^{-4}_{3}$ \\ \hline
		$k_{Y}=-2$ & $Y^{-2}_{1}$,~~$Y^{-2}_{3}$ \\ \hline
		$k_{Y}=0$ & $Y^{0}_{1}$,~~ $Y^{0}_{3}$  \\ \hline
		$k_{Y}=2$ & $Y^{2}_{1}$,~~$Y^{2}_{3}$  \\ \hline
		$k_{Y}=4$ & $Y^{4}_{1}$,~~$Y^{4}_{1^{\prime}}$~~$Y^{4}_{3}$   \\ \hline
		$k_{Y}=6$ & $Y^{6}_{1}$,~~$Y^{6}_{3I}$,~~$Y^{6}_{3II}$  \\ \hline
	\end{tabular}
	\caption{Polyharmonic $Maa\beta$ forms for different weight at level three}
	\label{w1t1}
\end{table}
In the present work, since we are working in the non-supersymmetric framework, we have focused on polyharmonic $Maa\beta$ forms of weight zero at level $3$. It can be arranged into a singlet and triplet under $A_{4}$ group. The q expansion of the weight zero Yukawa couplings at level three is provided below
	\begin{gather}\label{n4}
		\begin{aligned}
			Y^{(0)}_{3,1} &= y- \frac{3 e^{-4 \pi y}}{\pi q}- \frac{9 e^{-8\pi y}}{2\pi q^{2}}-\frac{-12\pi y}{\pi q^{3}}-\frac{21 e^{-16\pi y}}{4\pi q^{4}}-\frac{18 e^{-20\pi y}}{5\pi q^{5}}-\frac{3e^{-24\pi y}}{2\pi q^{6}}+\cdot\cdot\cdot\cdot\cdot \\& -\frac{9\log 3}{4\pi}-\frac{3q}{\pi}-\frac{9q^{2}}{2\pi}-\frac{q^{3}}{\pi}-\frac{21q^{4}}{4\pi}-\frac{18q^{5}}{5\pi}-\frac{3q^{6}}{2\pi} \\
			Y^{(0)}_{3,2} &= \frac{27q^{\frac{1}{3}}e^{\frac{\pi y}{3}}}{\pi}\big( \frac{e^{-3\pi y}}{4q} + \frac{e^{-7\pi y}}{5q^{2}} + \frac{5e^{-11\pi y}}{16q^{3}} + \frac{2e^{-15\pi y}}{11q^{4}} + \frac{2e^{-19\pi y}}{7q^{5}} + \frac{4e^{-23\pi y}}{17 q^{6}}+\cdot\cdot\cdot\cdot\cdot\cdot\cdot \big)\\& + \frac{9q^{\frac{1}{3}}}{2\pi}\big( 1+ \frac{7q}{4} + \frac{8q^{2}}{7}+ \frac{9q^{3}}{5}+\frac{14q^{4}}{13}+\frac{31q^{5}}{16}+\frac{20q^{6}}{19}+\cdot\cdot\cdot\cdot\big) \\
			Y^{(0)}_{3,3} &= \frac{9q^{\frac{2}{3}}e^{\frac{2\pi y}{3}}}{2\pi} \big(\frac{e^{-2\pi y}}{q}+\frac{7e^{-6\pi y}}{4q^{2}}+\frac{8e^{-10\pi y}}{7q^{3}}+ \frac{9e^{-14\pi y}}{5q^{4}}+\frac{14e^{-18\pi y}}{13q^{5}}+\frac{31e^{-22\pi y}}{16q^{6}}+\cdot\cdot\cdot\cdot\cdot\cdot \big) \\& +\frac{27q^{\frac{2}{3}}}{\pi}\big(\frac{1}{4}+\frac{q}{5}+\frac{5q^{2}}{16}+\frac{2q^{3}}{11}+\frac{2q^{4}}{7}+\frac{9q^{5}}{17}+\frac{21q^{6}}{20}+\cdot\cdot\cdot\cdot\cdot \big)
		\end{aligned}    
\end{gather}
$A_{4}$ is a finite modular group and isomorphic to $\Gamma_{3}$ group. $A_{4}$ group is a non-Abelian discrete group and has a total of four irreducible representations, among which three are one-dimensional and one is three-dimensional. A brief introduction about the $A_{4}$ group is given in the Appendix \textbf{C}.\\
\section{\textbf{The Model}} \label{s4}
To implement $\Gamma_{3}$ modular group, we have considered the intermediate asymmetric gauge group $SU(2)_{L}\times U(1)_{R} \times U(1)_{B-L}\times SU(3)_{C}$ and have used the scalar triplet ($\Delta_{R}$) and doublet ($\chi_{R}$) to break the intermediate asymmetric group to the SM gauge group. We have considered an extra sterile fermion per generation to generate the light neutrino mass via extended inverse seesaw mechanism \cite{Senapati:2020alx,Awasthi:2013ff,Parida:2012sq}.The Yukawa  Lagrangian term associated with the model is given in equation \ref{Q4}
\begin{equation}\label{Q4}
	\mathcal{L} = \mathcal{L}_{l} + \mathcal{L}_{D} +\mathcal{L}_{M} +\mathcal{L}_{N-S} +\mathcal{L}_{S} + h.c ~~.
\end{equation}
Where $\mathcal{L}_{l}$ is the Yukawa Lagrangian term for the charged lepton, $\mathcal{L}_{D}$ is the Dirac Yukawa Lagrangian term for the neutral lepton, $\mathcal{L}_{M}$ and $\mathcal{L}_{N-S}$ are the Majorana and neutrino-sterile (N-S) mixing terms respectively and finally $\mathcal{L}_{S}$ is the self-coupling term for the sterile fermion. In Table \ref{TAB2} we have provided the charge assignments and modular weights for the particle contents of the model. The charge assignment under $U(1)_{R}$ corresponds to the third or z-component of isospin i.e., $T_{3R}$ of $SU(2)_{R}$ and the equation $ Q = T_{3L} + T_{3R} + \frac{B-L}{2}$ is  also valid for the gauge group $G_{2113}$.
\begin{table}[ht]
	\centering
	\begin{tabular}{|m{1.5cm}|c| c|c|c|c|c|c|}
		\hline
		
		Field & $\Psi_{R_{i}}$ &$ \Psi_{L_{i}} $ & $N_{R} $ & $  S$ & $\Phi$ & $\Delta_{R}$ & $\chi_{R}$  \\ \hline
		
		$ SU(2)_{L}$& 1& 2 & 1 & 1 & 2 & 1 & 1 \\ \hline
		
		$U(1)_R$ & $-\frac{1}{2}$ &0 & $\frac{1}{2}$ & 0 &$-\frac{1}{2}$ & -1 &$\frac{1}{2}$ \\ \hline
	
		$ A_{4} $ & 1,$1^{\prime\prime}$,$1^{\prime}$ & 1,$1^{\prime}$,$1^{\prime\prime}$ & 3 & 3 & 1 & 1 & 1 \\ \hline
		
		$k_{I}$ & 0 & $ 0 $ & 0 & 0 & 0 & 0 & 0 \\ \hline
	\end{tabular}
	\caption{Charge assignment for the particle content of the model}
	\label{TAB2}
\end{table}
\subsection{\textbf{Mass term for charged leptons}} 
To construct a diagonal charged lepton mass matrix, we consider the three-generation left-handed lepton doublet $\Psi_{L_{i}}(i=1,2,3)$ transforming as $1,1^{\prime},1^{\prime\prime}$ under the $A_{4}$ group. Similarly, the three right-handed charged leptons $\Psi_{R_{i}}(i=1,2,3)$ transform as $1,1^{\prime\prime},1^{\prime}$ under the $A_{4}$ group. The Yukawa Lagrangian term for the charged leptons and the corresponding diagonal charged lepton mass matrix are given in equations \eqref{w1q1} and \eqref{w1q2}, respectively. \\  
\begin{equation}\label{w1q1}  
	\mathcal{L}_{l} = \alpha \Phi \bar{\Psi}_{L_{1}} Y^{0}_{1}\Psi_{R_{1}} + \beta \Phi \bar{\Psi}_{L_{2}} Y^{0}_{1}\Psi_{R_{2}}+\gamma \Phi \bar{\Psi}_{L_{3}} Y^{0}_{1}\Psi_{R_{3}}  
\end{equation} 
The modular form $Y^{0}_{1}$ is a constant, while $\alpha$, $\beta$, and $\gamma$ are adjustable parameters. By choosing appropriate values for these parameters, we obtain the desired charged lepton mass matrix.  
\begin{equation}\label{w1q2}  
	M_{D}= v \begin{pmatrix}  
		Y^{0}_{1}\alpha & 0 & 0 \\  
		0 & Y^{0}_{1}\beta & 0 \\  
		0 & 0 & Y^{0}_{1}\gamma  
	\end{pmatrix} ~~ .  
\end{equation}
\subsection{\textbf{Dirac mass term for neutrino}}
To construct the $A_4$ invariant Dirac mass term, we assume that the lepton doublets $\Psi_{L_{i}}$ transform as $1,1^{\prime},1^{\prime\prime}$, while the right-handed neutrinos $N_{R}$ transform as triplets, and the Higgs bidoublet transforms as a singlet under the $A_4$ group. The corresponding $A_4$ invariant Dirac Yukawa interaction is given in \eqref{Q7}.
\begin{equation}\label{Q7}
	\mathcal{L}_{D} =g_{1}\Phi \bar{\Psi}_{L_{1}}(Y^{(0)}_{3}N_{R})_{1}+g_{2}\Phi \bar{\Psi}_{L_{2}}(Y^{(0)}_{3}N_{R})_{1^{\prime\prime}}+g_{3}\Phi \bar{\Psi}_{L_{3}}(Y^{(0)}_{3}N_{R})_{1^{\prime}}~~.
\end{equation} 
Where $g_{1}$, $g_{2}$ and $g_{3}$ are adjustable parameters of the model. From equation \eqref{Q7}, we can construct the Dirac mass matrix of the model, which is given in equation \eqref{Q8}
\begin{equation}\label{Q8}
	M_{D}= v \begin{pmatrix}
		g_{1} Y^{(0)}_{3,1} & g_{1}Y^{(0)}_{3,3} & g_{1} Y^{(0)}_{3,2} \\
		g_{2}Y^{(0)}_{3,3} & g_{2}Y^{(0)}_{3,2} & g_{2}Y^{(0)}_{3,1} \\
		g_{3}Y^{(0)}_{3,2} & g_{3}) Y^{(0)}_{3,1} & g_{3}Y^{(0)}_{3,3}
	\end{pmatrix} ~~ .
\end{equation}
Where $v=246$ GeV is the V.E.V of the bidoublet $\Phi$, and ($g_{1}$, $g_{2}$,$g_{3}$) are adjustable complex parameters of the model. Their values are chosen such that all the neutrino oscillation parameters fall within the $3\sigma$ range and the real and imaginary parts of those free parameters lie within the range of 0.4 to 0.8.
\subsection{\textbf{Majorana mass term for neutrino}}
We have  assigned a singlet to the scalar triplet $\Delta_{R}$, and  by considering the symmetric nature of the Majorana mass term we can write the Majorana mass term, in the following way:
\begin{equation}\label{Q9}
	\mathcal{L}_{M_{R}} = g_{4}\Delta_{R} Y^{(0)}_{3}(\bar{N^{C}}_{R}N_{R})_{3_S} +  g_{5}\Delta_{R} Y^{(0)}_{1}(\bar{N^{C}}_{R}N_{R})_{1} ~~ .
\end{equation}
From equation \eqref{Q9}, we have constructed the Majorana mass matrix. In equation \eqref{Q10}, $v_{R}$ is the V.E.V of the scalar triplet $\Delta_{R}$, and is taken to be $3$~TeV in our analysis. $g_{4}$ and $g_{5}$ are complex adjustable parameters of the model and we have chosen their real and imaginary parts to lie within the range of 0.4 to 0.8.
\begin{equation}\label{Q10}
	M=v_{R} \begin{pmatrix}
		g_{5}+2 g_{4} Y^{(0)}_{3,1} & -g_{4}Y^{(0)}_{3,3} & - g_{4}Y^{(0)}_{3,2} \\
		-g_{4}Y^{(0)}_{3,3}& 2g_{4} Y^{(0)}_{3,2} & g_{5}-g_{4}Y^{(0)}_{3,1} \\
		-g_{4}Y^{(0)}_{3,2} & g_{5}-g_{4}Y^{(0)}_{3,1} & 2 g_{4}Y^{(0)}_{3,3}
	\end{pmatrix}~~
\end{equation}
\subsection{\textbf{N-S mixing term}}
The sterile fermion $S_{i}~(i=1, 2, 3)$ and the scalar doublet $\chi_{R}$  both transform as singlets under the $A_{4}$ group, and the scalar doublet $\chi_{R}$ is responsible for $N-S$ mixing. The $A_{4}$ invariant $N-S$ mixing term is given in the equation \eqref{Q11}
\begin{equation}\label{Q11}
	\mathcal{L}_{N} = f_{1}\chi_{R}(\bar{N}_{R}S)_{3_{S}} Y^{(0)}_{3}+ f_{2}\chi_{R}(\bar{N}_{R}S)_{3_{A}}Y^{(0)}_{3} + f_{3}\chi_{R}(\bar{N}_{R}S)_{1}Y^{0}_{1} ~~ .
\end{equation} 
Here, $f_{1}$, $f_{2}$ and $f_{3}$ are complex, adjustable parameters of the model, whose values can be adjusted to achieve the desired results. In the present case, their real and imaginary components are assumed to lie within the range of $0.1$ to $1$. From the above equation, we have constructed the $N-S$ mixing mass matrix, which is given in equation \eqref{Q12}
\begin{equation}\label{Q12}
		M = v^{\prime} \begin{pmatrix}
			f_{3}+2f_{1} Y^{(0)}_{3,1} & (-f_{1}+f_{2})Y^{(0)}_{3,3} & (-f_{1}-f_{2})Y^{(0)}_{3,2} \\
			(-f_{1}-f_{2})Y^{(0)}_{3,3} & 2f_{1}Y^{(0)}_{3,2} & f_{3}+(-f_{1}+f_{2})Y^{(0)}_{3,1} \\
			(-f_{1}+f_{2})Y^{(0)}_{3,2} & f_{3}+(-f_{1}-f_{2}) Y^{(0)}_{3,1} & 2f_{1}Y^{(0)}_{3,3}
		\end{pmatrix} ~~ .
\end{equation}

The V.E.V associated with the scalar doublet $\chi_{R}$, represented by $v^{\prime}$, is assumed to be $1$~TeV in the present analysis.
\subsection{\textbf{Sterile-Sterile ($S-S$) mixing term}}
This is the Majorana mass term for the sterile fermion. Equations \eqref{Q13} and \eqref{Q14}, represent the $S-S$ mixing mass term and the corresponding mass matrix, respectively. We have taken $g_{6}$ and $g_{7}$ within the range of 10 keV to 50 keV.
\begin{equation}\label{Q13}
		\mathcal{L}_S= g_{6} (S^{T}S) + g_{7}Y^{0}_{3}(S^{T}S)_{3_{S}}~~.
\end{equation}
\begin{equation}\label{Q14}
	M_{S}=\begin{pmatrix}
			g_{6}+2 g_{7} Y^{(0)}_{3,1} & -g_{7}Y^{(0)}_{3,3} & - g_{7}Y^{(0)}_{3,2} \\
		-g_{7}Y^{(0)}_{3,3}& 2g_{7} Y^{(0)}_{3,2} & g_{6}-g_{7}Y^{(0)}_{3,1} \\
		-g_{7}Y^{(0)}_{3,2} & g_{6}-g_{7}Y^{(0)}_{3,1} & 2 g_{7}Y^{(0)}_{3,3}
	\end{pmatrix}~~.
\end{equation}
\section{\textbf{Extended inverse seesaw mechanism}} \label{s5}
Spontaneous symmetry breaking of the Yukawa Lagrangian term given in equation \eqref{Q4} gives rise to a $9\times9$ mass matrix, which is given in equation \ref{Q15} 
\begin{equation}\label{Q15}
	\begin{pmatrix}
		0 & 0 & M_{D} \\
		0 & M_{S} & M \\
		M^{T}_{D} & M^{T} & M_{R}
	\end{pmatrix} ~~.
\end{equation}  
By considering the mass hierarchy  scale as $M_{R}> M>>M_{D}$ we can block diagonalize this $9\times 9$ matrix, and we  will finally obtain the mass matrices for the light neutrino, sterile fermion, and right handed neutrino which is given respectively in equations \eqref{Q16} . 
\begin{gather} \label{Q16}
	\begin{aligned}
		m_{\nu} & = M_{D}M^{-1}M_{S}(M_{D}M^{-1})^{T}  \\ 
		m_{S}  &= M_{S} - MM^{-1}_{R}M^{T}  \\ 
		m_{R} &= M_{R}~ .
	\end{aligned}
\end{gather}
To obtain the eigenvalues we further diagonalized the  matrices given in equation \eqref{Q16} by their respective unitary matrices as follows
\begin{gather} \label{q17}
	\begin{aligned}
		\hat{m}_{\nu} & = U^{\dagger}_{\nu}m_{\nu}U_{\nu}^{*} = diag(m_{\nu_{1}}, m_{\nu_{2}}, m_{\nu_{3}})\\
		\hat{m}_{S} & =U^{\dagger}_{S}m_{s}U^{*}_{S} = diag(m_{S_{1}}, m_{S_{2}}, m_{S_{3}}) \\
		\hat{m}_{R} & = U^{\dagger}_{N}m_{R}U^{*}_{N} = diag(m_{R_{1}}, m_{R_{2}}, m_{R_{3}})~~~. \\
	\end{aligned}
\end{gather}
The complete mixing matrix responsible for diagonalizing the $9\time9$ mass matrix given in the equation  \ref{Q15} is shown bellow \cite{Senapati:2020alx,Awasthi:2013ff,Grimus:2000vj}
\begin{align}\label{Q18}
	\mathbf{V}=\begin{pmatrix}
		V^{\nu\nu}& V^{\nu S} & V^{\nu N} \\
		V^{S\nu} & V^{SS} & V^{SN} \\
		V^{N\nu} & V^{NS} & V^{NN} 
	\end{pmatrix}=
	\begin{pmatrix}
		(1-\frac{1}{2}XX^{\dagger})U_{\nu} & (X-\frac{1}{2}ZY^{\dagger})U_{S}& ZU_{N} \\
		-XU_{\nu} & (1-\frac{1}{2}(X^{\dagger}X+YY^{\dagger}))U_{S} & (Y-\frac{1}{2}X^{\dagger}Z)U_{N}\\
		y^{*}X^{\dagger}U_{\nu} & -Y^{\dagger}U_{S} & (1-\frac{1}{2}Y^{\dagger}Y)U_{N}
	\end{pmatrix} 
\end{align}
where $X=M_{D}M^{-1},~ Y=MM_{N}^{-1},~Z=M_{D}M^{-1}_{N},~$ and $y=M^{-1}M_{S}$ \\
Where $U_{\nu}=V_{l}^{\dagger}V_{\nu}$, but in this model we have considered the charged lepton mass basis is diagonal i.e. $V_{l}$ is an identity matrix so $V_{\nu}=(1-\eta)U_{PMNS}$ and the parameter $\eta$ represents the deviation from unitarity. The matrix $U_{PMNS}$ is a unitary matrix and it can be parametrized by using three mixing angles and three phases, among which one is a Dirac CP phase denoted as $\delta_{CP}$ and two are Majorana phases ($\alpha$, $\beta$). By using standard parametrization, we can write the $U_{PMNS}$ matrix in the following way
\begin{align}\label{Q19}
	U_{PMNS}= \begin{pmatrix}
		c_{12}c_{13} & s_{12}c_{13} & s_{13}e^{-i\delta_{CP}} \\
		-s_{12}c_{23}-c_{12}s_{23}s_{13}e^{i\delta_{CP}} & c_{12}c_{23}-s_{12}s_{23}s_{13}e^{i\delta_{CP}} & s_{23}c_{13} \\
		s_{12}s_{23}-c_{12}c_{23}s_{13} e^{i\delta_{CP}} & -c_{12}s_{23}-s_{12}c_{23}s_{13}e^{i\delta_{CP}} & c_{23}c_{13} \\
	\end{pmatrix}
	\begin{pmatrix}
		1 & 0 & 0\\
		0 & e^{i\alpha} & 0\\
		0 & 0 & e^{i\beta}
	\end{pmatrix}~.
\end{align}
Where $c_{ij}$ and $s_{ij}$ stands for $\cos\theta_{ij}$ and $\sin\theta_{ij}$, respectively. One can represent all the mixing angles in terms of the elements of the $U_{PMNS}$ matrix as given in equation \eqref{Q20}, and their $3\sigma$ values of neutrino oscillation parameters are given in the Table \ref{TAB4}. We have used those $ 3\sigma$ ranges for our numerical analysis part.
\begin{align}\label{Q20}
	\sin^{2}\theta_{13} = |(U_{PMNS})_{13}|^{2} ~, ~~ \sin^{2}\theta_{23} = \frac{|(U_{PMNS})_{23}|^{2}}{1-|(U_{PMNS})_{13}|^{2}} ~, ~~ \sin^{2}\theta_{12}=\frac{|(U_{PMNS})_{12}|^{2}}{1-|(U_{PMNS})_{13}|^{2}}
\end{align}
The Dirac CP phase, Jarlskog invariant and Majorana phases can also be estimated from $U_{PMNS}$ matrix, and their relations are given in the equation \eqref{Q21} and \eqref{Q22} respectively

\begin{equation}\label{Q21}
	J_{CP}= Im[U_{e1}U_{\mu2}U_{e2}^{*}U_{\mu1}^{*}] = s_{23}c_{23}s_{12}c_{12}s_{13}c^{2}_{13}\sin\delta_{CP} 
\end{equation}
\begin{equation}\label{Q22}
	Im[U_{e1}^{*}U_{e2}]=c_{12}s_{12}c^{2}_{13}\sin\alpha ~, ~~ Im[U_{e1}^{*}U_{e3}]=c_{12}s_{13}c_{13}\sin(\beta-\delta_{CP}). 
\end{equation}
\begin{table}[h]
	\centering
	\begin{tabular}{|c|c|c|} \hline
		\textbf{Oscillation Parameter} & \textbf{Normal Hierarchy} & \textbf{Inverted Hierarchy} \\ \hline
		
	$\sin^{2}\theta_{12}$ & $0.275 \rightarrow 0.345$ & $0.275 \rightarrow 0.345$ \\ [4ex] \hline
	$\sin^{2}\theta_{23}$ & $0.435 \rightarrow 0.585$ & $0.440 \rightarrow 0.585$ \\ [4ex] \hline
	$\sin^{2}\theta_{13}$ & $0.02030 \rightarrow 0.02388$ & $0.02060 \rightarrow 0.02409$ \\ [4ex] \hline
	$\delta_{CP}$ & $124 \rightarrow 364$ & $201 \rightarrow 335$ \\ [4ex] \hline
	$\frac{\Delta m_{21}^{2}}{10^{-5}}$ & $6.92 \rightarrow 8.05$ & $6.92 \rightarrow 8.05$ \\ [4ex] \hline
	$\frac{\Delta m_{3l}^{2}}{10^{-3}}$ & $+2.451 \rightarrow 2.578$ & $-2.547 \rightarrow -2.421$ \\ [4ex] \hline
	\end{tabular}
	\caption{$3\sigma$ values of oscillation parameters~\cite{Esteban:2024eli}}
	\label{TAB4}
\end{table}
\section{Non-unitary effect} \label{s6}
The PMNS matrix is responsible for diagonalizing the light neutrino mass matrix, but in the case of the extended inverse seesaw mechanism, due to the mixing between light and heavy neutrinos even after the diagonalization of the light neutrino mass matrix, we get off-diagonal terms. The complete $9\times9$ diagonalizing matrix given in the equation \eqref{Q18} can be decomposed in the following way \cite{Dev:2009aw,LalAwasthi:2011aa}
\begin{equation}\label{Q23}
	\mathbf{V} = \begin{pmatrix}
		\mathbf{V}_{3\times3} &\mathbf{V}_{3\times6}\\
		\mathbf{V}_{6\times3} & \mathbf{V}_{6\times6}
	\end{pmatrix}~~.
\end{equation}
The PMNS matrix $\mathbf{V}_{3\times3}=(1-\eta)U_{\nu}$ is responsible for diagonalizing the light neutrino mass matrix, and it is not a unitary matrix. The $3\times3$ Hermitian matrix $\eta= \frac{1}{2}XX^{\dag}$ is known as the deviation from unitarity, which only depends on the Dirac and Majorana matrices of the model. Upper bounds on the elements of this Hermitian matrix \cite{Fernandez-Martinez:2016lgt} have been determined from various experiments, and the $2\sigma$  constraints on these values are given below

\begin{equation}\label{QQ1}
	\eta \leq \begin{pmatrix}
		2.5\times 10^{-3} & 2.4 \times 10^{-5} & 2.7 \times 10^{-3} \\
		2.4 \times 10^{-5} & 4.0 \times 10^{-4} & 1.2 \times 10^{-3} \\
		2.7 \times 10^{-3} & 1.2 \times 10^{-3} & 5.6 \times 10^{-3}
	\end{pmatrix}
\end{equation}
Due to this non-unitary matrix, the Jarlskog invariant term gets modified, and it can be written in the following way
\begin{equation}\label{Q24}
	J^{ij}_{\alpha\beta}= Im(\mathbf{V}_{\alpha i}\mathbf{V}_{\beta j}\mathbf{V}^{*}_{\alpha i}\mathbf{V}^{*}_{\beta j}) \approx J_{CP} + \Delta J^{ij}_{\alpha\beta}
\end{equation}
where $\alpha \neq \beta$ and they represent the charged leptons ($e, \mu, \tau$), $i \neq j$ and can take the value $1, 2, 3$. The first term in the equation \eqref{Q24}, represents the effect on CP-violation due to the unitary PMNS matrix and the second term represents the contribution to CP-violation due to non-unitary term.
\begin{equation*}
	\Delta^{ij}_{\alpha\beta} \approx -\sum_{\gamma=e,\mu,\tau} Im(\eta_{\alpha\gamma}U_{\alpha i}U_{\beta j}U_{\alpha j}^{*}U^{*}_{\beta i} + \eta_{\beta\gamma}U_{\alpha i}U_{\gamma j}U_{\alpha j}^{*}U_{\beta i}^{*}+\eta_{\alpha\gamma}^{*}U_{\alpha i}U_{\beta j}U_{\gamma j}^{*}U_{\beta i}^{*} + \eta_{\beta\gamma}^{*}U_{\alpha i}U_{\beta j}U_{\alpha j}^{*}U_{\gamma i}^{*}) ~~.
\end{equation*} 
Usually, $J_{CP}$ becomes equal to zero when $\delta_{CP}$ and $sin\theta_{13}$ tend to zero but when we consider non-unitary effects Jarlskog invariant term remains non-zero even if $\delta_{CP}$ and $\sin\delta_{CP}$ become zero.   
\section{\textbf{Neutrinoless double beta decay}} \label{s7}
Neutrinoless double beta ($0\nu\beta\beta$) is a very fascinating phenomenon in which a nucleus undergoes two successive beta decays without emitting any neutrinos in the final process. This hypothetical nuclear decay process has not yet been observed. It is a type of beta decay in which two neutrons in the nucleus simultaneously convert into two protons, emitting two electrons but no accompanying neutrinos. The process can be denoted in the following way
\begin{equation*}
	N(A,Z) \rightarrow N(A, Z+2) + 2e^{-}
\end{equation*}
This process violates the conservation of the lepton number, which is an accidental symmetry of the Standard Model. The observation of such decay would imply that neutrinos are their own antiparticles and have a Majorana nature, meaning that the neutrino and antineutrino are indistinguishable. There are several processes that contribute to the $0\nu\beta\beta$ decay in this model. These processes are discussed below
\subsubsection{Contribution due to $W_{L}-W_{L}$ current}
In the case of $W_{L}-W_{L}$ mediation, there are three possible contributions and among them, popular contribution to $0\nu\beta\beta$ decay is the standard contribution i.e. contribution due to the exchange of light Majorana neutrino and the other two contributions are due to the exchange of right-handed and sterile neutrino, which is also known as the non-standard contribution.
\subsubsection{Contribution due to $W_{R}-W_{R}$ current}
In case when the mediator is purely right-handed current ($W_{R}-W_{R}$), there are three potential contributions to neutrinoless double beta decay ($0\nu\beta\beta$). These contributions arise from the exchange of light neutrinos, right-handed, and sterile neutrinos. 
\subsubsection{Contribution due to $W_{L}-W_{R}$ current }
In the case of $W_{L}-W_{R}$ mediation, we can have two types of mixed helicity Feynman diagram, which is known as the $\lambda$ and $\eta$ diagram.
\subsubsection{Contribution due to charged Higgs boson}
The left and right-handed doubly charged Higgs boson namely $\Delta_{L,R}$ also contribute to the $0\nu\beta\beta$ decay but its contribution is negligible as compared to other processes. \\
The Feynman diagram and amplitude of all the above mention processes are given in the Appendix \ref{AA}.
\section{\textbf{Numerical Analysis and Results}} \label{s8}
In case of Extended inverse seesaw mechanism the light neutrino mass matrix is given by
\begin{equation}\label{qq1}
m_{\nu}=M_{D}M^{-1}M_{S}(M_{D}M^{-1})^{T}
\end{equation}
We can also write light neutrino mass matrix in terms of $U_{PMNS}$ matrix in the following way
\begin{equation}\label{qq2}
m_{\nu}=(1-\eta)U_{PMNS}~m_{dig}~((1-\eta)U_{PMNS})^{\dagger}
\end{equation}
After constructing the mass matrices utilizing the multiplication rules of the $A_{4}$ group, we incorporated these matrices into equation \eqref{qq1} to derive the light neutrino mass matrix in terms of Yukawa couplings. To further analyze this, we computed the values of the Yukawa couplings using the $3\sigma$ values of oscillation parameters, which were then inserted into equation \eqref{qq2}. Subsequently, we equated equations \eqref{qq1} and \eqref{qq2} to determine the unknown values of the Yukawa couplings. Table \ref{TAB10} provides the Yukawa coupling values.

\begin{table}[ht]
	\centering
	\begin{tabular}{|c|c|c|}
		\hline
		Yukawa  coupling & for NH & for IH \\ \hline
	$Y^{(0)}_{3,1}$ & $0.00069-6.725$ & $0.00171-8.829$ \\ \hline
	$Y^{(0)}_{3,2}$ & $0.00098-6.362$ & $0.00090-6.783$ \\ \hline
	$Y^{(0)}_{3,3}$ & $0.00025-6.226$ & $0.00041-6.700$\\ \hline
	\end{tabular}
	\caption{Yukawa coupling values for NH and IH}
	\label{TAB10}
\end{table}	
\begin{figure}[h]
	\centering
	\begin{subfigure}[b]{0.45\textwidth}
		\centering
		\includegraphics[width=7cm, height=6cm]{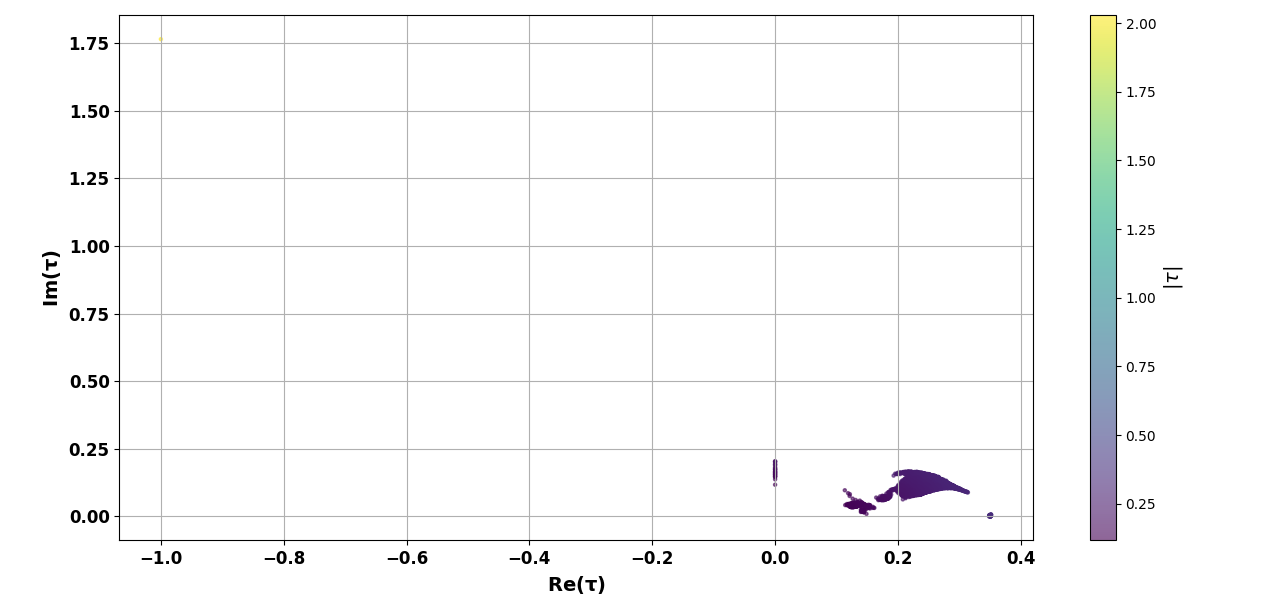}
		\caption{Variation of Re$\tau$ and Im$\tau$}
		\label{f5}
	\end{subfigure}
	\hfill
	\begin{subfigure}[b]{0.45\textwidth}
		\centering
		\includegraphics[width=7cm, height=6cm]{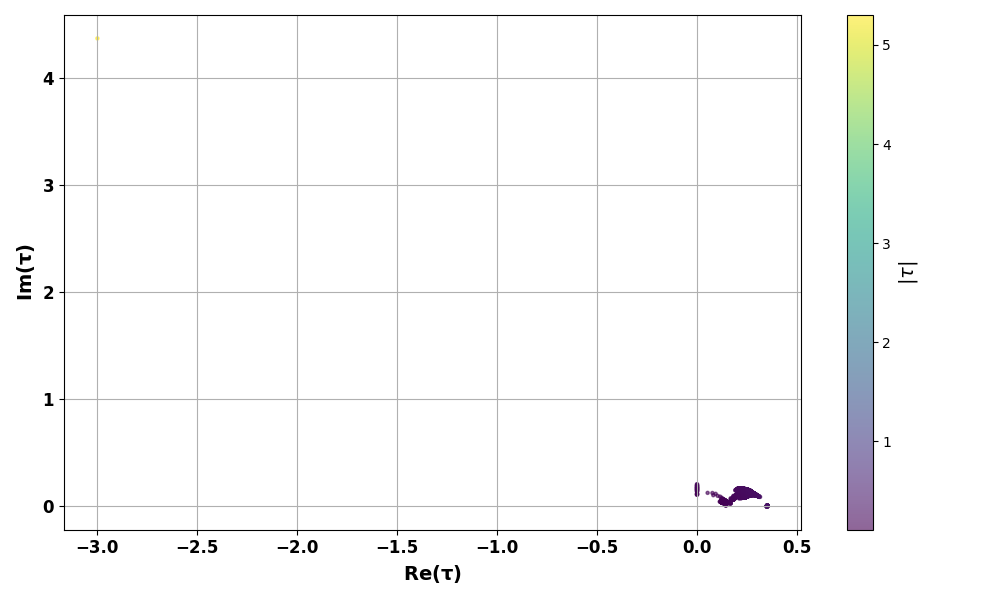}
		\caption{Variation of Re$\tau$ and Im$\tau$}
		\label{f6}
	\end{subfigure}
	\caption{The two figures show the parameter space for the real and imaginary parts of the modulus $\tau$ for NH and IH, respectively.}
	\label{fig2}
\end{figure}
The motivation behind utilizing the modular group is to minimize the necessity of introducing an additional field, commonly known as a "flavon", for breaking the discrete flavor symmetry. Within modular symmetry, the modulus $\tau$ is responsible for breaking the  discrete flavor symmetry. We calculated the values of the modulus $\tau$ by employing the $q$ expansion of the modular form as given in equation \eqref{n4}. We have computed the absolute value of the modulus $\tau$, as well as its real and imaginary components, from our model. For the NH case, the real part of $\tau$ ranges from $-0.999$ to $0.35$, and the imaginary part ranges from $2.8 \times 10^{-5}$ to $1.764$. However, as shown in Figure \ref{f5}, we observe that most of the real part values of $\tau$ are concentrated within the range $0.1$ to $0.3$, while the imaginary part predominantly spans from $2.8 \times 10^{-5}$ to below $0.25$. Similarly, for the IH case, the computed real part of $\tau$ lies within the range $-3.0$ to $0.35$, and the imaginary part spans $1.5\times10^{-5}$ to $4.37$. However, Figure \ref{f6} indicates that the real part of $\tau$ is primarily clustered between $0.1$ and $0.4$, while the imaginary part mostly occupies the range $1.5\times10^{-5}$ to $0.2$. Additionally, Figure \ref{fig2} provides insight into the parameter space of the absolute value of $\tau$. We find that, in the case of NH, the modulus $|\tau|$ ranges from $0.117$ to $2.028$, whereas for IH, it spans from $0.109$ to $5.302$.
\subsection{Active neutrino masses}
\begin{figure}[h]
	\begin{subfigure}[b]{0.45\textwidth}
		\centering
		\includegraphics[width=\linewidth]{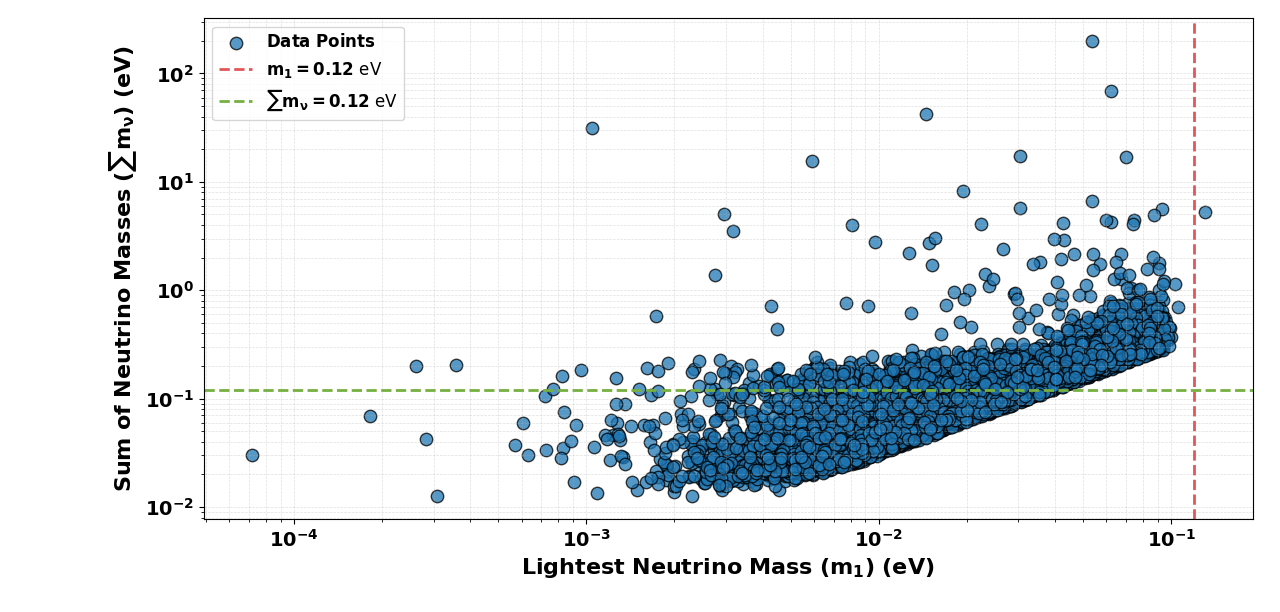}
		\caption{For NH}
		\label{f9}
	\end{subfigure}
	\hfill
	\begin{subfigure}[b]{0.45\textwidth}
		\centering
		\includegraphics[width=\linewidth]{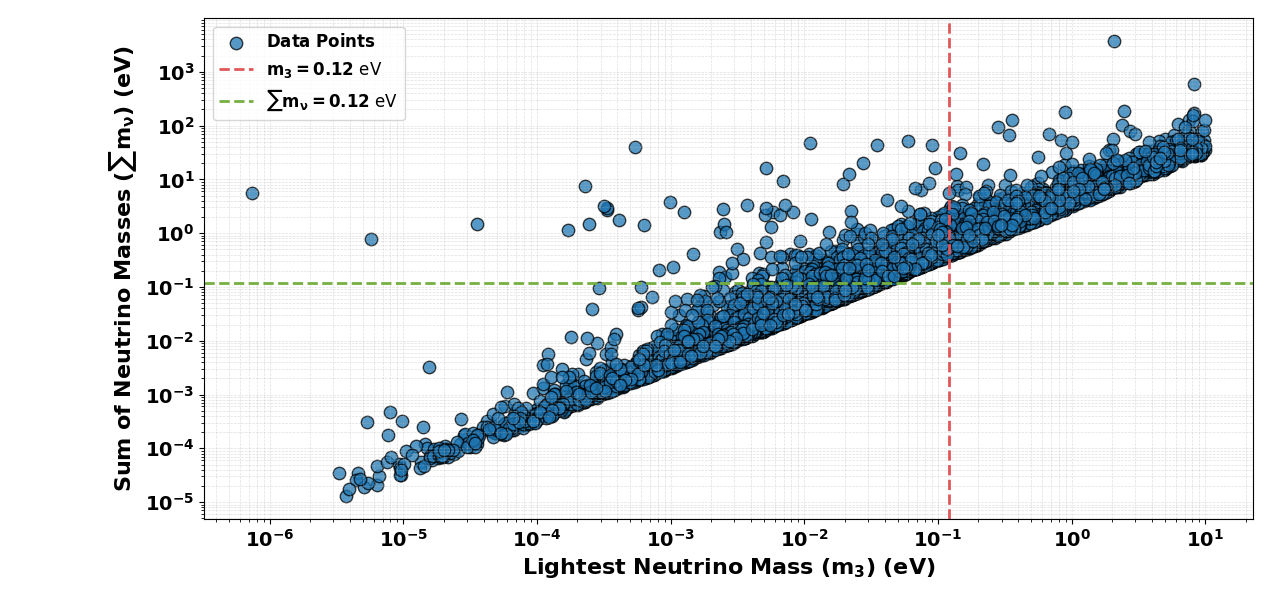}
		\caption{For IH}
		\label{f10}
	\end{subfigure}
	\hfill
 \begin{subfigure}[b]{0.45\textwidth}
		\centering
		\includegraphics[width=\linewidth]{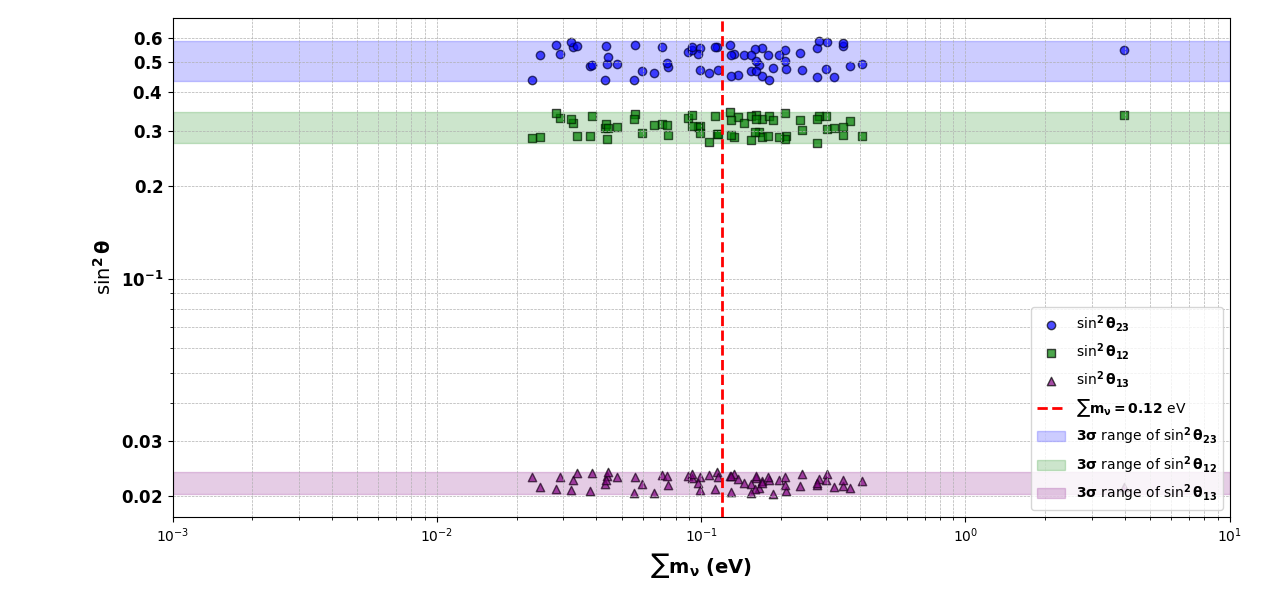}
		\caption{For NH}
		\label{f13}
	\end{subfigure} 
	\hfill
	\begin{subfigure}[b]{0.45\textwidth}
		\centering
		\includegraphics[width=\linewidth]{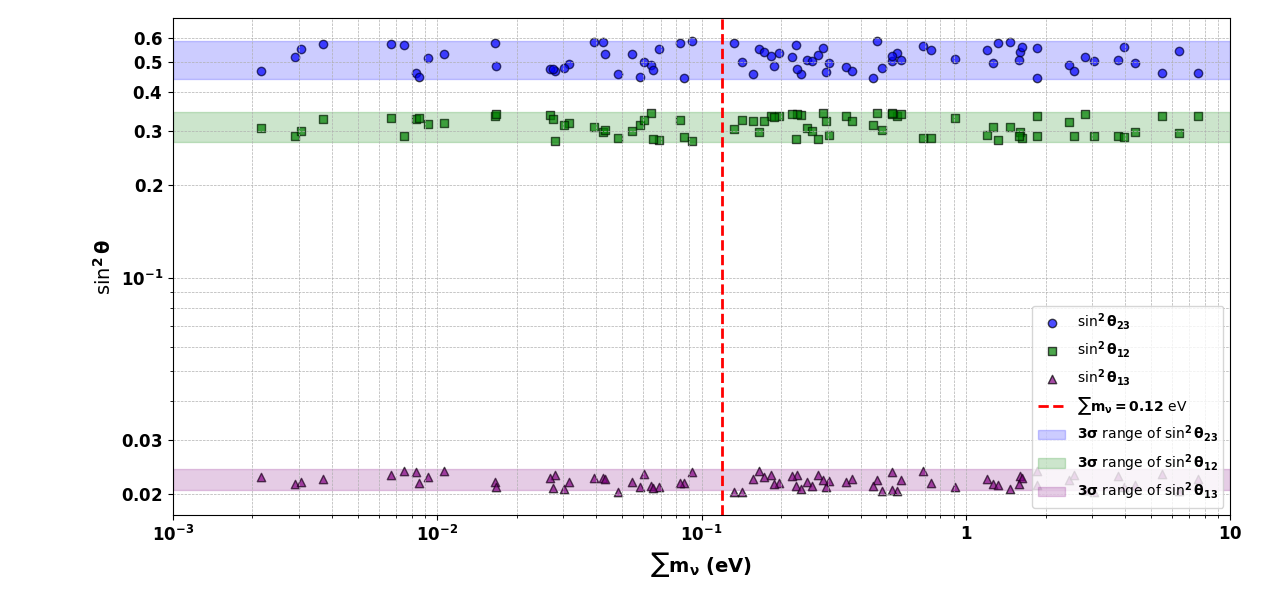}
		\caption{For IH}
		\label{f15}
	\end{subfigure}
	\caption{The top two Figures show variations of sum of the neutrino masses with the lightest neutrino mass for NH and IH and the bottom two Figures show the variation of mixing angles with the sum of the neutrino masses for NH and IH.}
	
	\label{fig3}
\end{figure}

\begin{figure}[h]
	\begin{subfigure}[b]{0.45\textwidth}
		\centering
		\includegraphics[width=\linewidth]{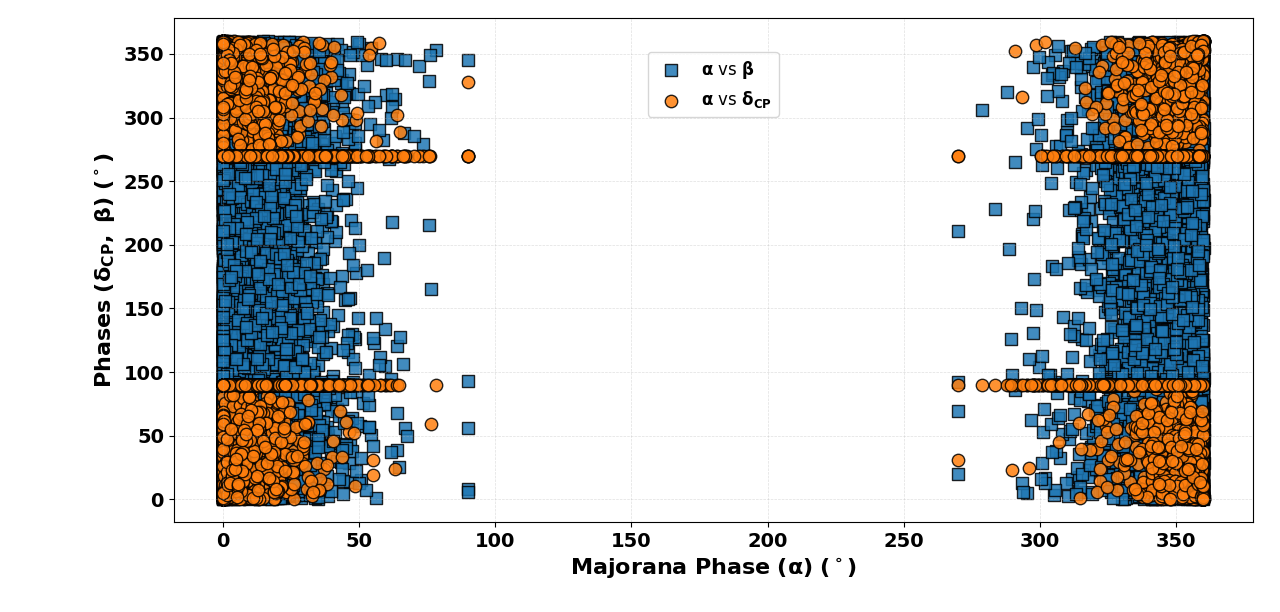}
		\caption{For NH}
		\label{f11}
	\end{subfigure} 
	\hfill
	\begin{subfigure}[b]{0.45\textwidth}
		\centering
		\includegraphics[width=\linewidth]{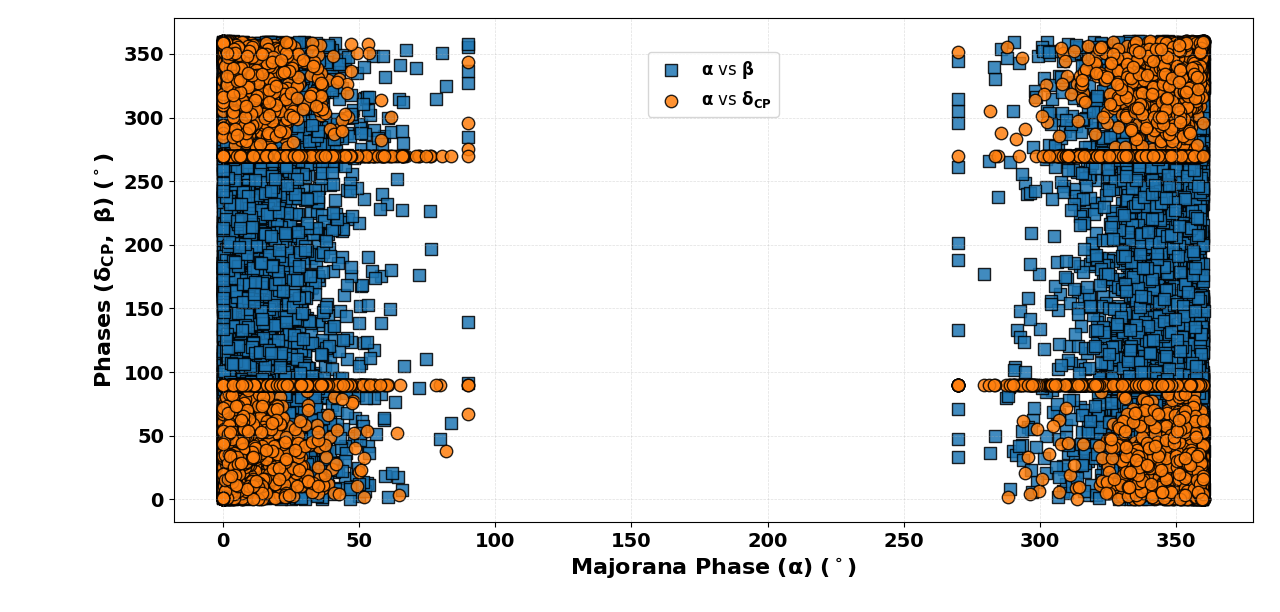}
		\caption{For IH}
		\label{f12}
	\end{subfigure} 
	\caption{Figures show the parameter space of $\delta_{CP}$ and Majorana phases ($\alpha$ and $\beta$) for NH and IH. }
	
	\label{fig4}
\end{figure}
We have plotted a graph depicting the relationship between the sum of neutrino masses and the lightest neutrino mass, considering both the normal and inverted hierarchies. It appears that the majority of the calculated values lie below the experimental bound. This experimental bound is represented by a horizontal and vertical lines in Figures ~\ref{f9} and \ref{f10}. To verify the consistency of the model, we have also plotted the mixing angles calculated from the model against the sum of the neutrino masses for both NH and IH. Figures~ \ref{f13} and \ref{f15} show the variation of the three mixing angles with the sum of the neutrino masses for NH and IH respectively. We have observed that most of the data fall within the experimental $3\sigma$ range. We have also calculated the value of the two Majorana phases and $\delta_{CP}$ from our model. From Figures ~\ref{f11} and \ref{f12}, it is observed that the Majorana phase $\alpha$ lies within the range of [$0^{\circ}-50^{\circ}$] and [$300^{\circ}-350^{\circ}$], moreover  $\delta_{CP}$ values are same as those of the Majorana phase $\alpha$. Majorana phase $\beta$ covers all values from $0^{\circ}$ to $350^{\circ}$ and it holds true for both the NH and IH cases.
\subsection{Heavy neutrino Eigenstates}
Within this model, we have a total of six heavy mass eigenstates, comprising three right-handed neutrinos and three sterile fermions. To analyze the parameter space for right-handed neutrino masses, we have plotted the variation of $M_{N_{1}}$ against $M_{N_{2}}$ and $M_{N_{3}}$ for both NH and IH. The model predicts that the right-handed neutrino masses range from $1$ TeV to $100$ TeV. Similarly, we have studied the parameter space for the masses of sterile fermions. As shown in Figures~ \ref{f23} and \ref{f24}, the sterile fermion masses range from $0.01$ TeV to approximately $10^{4}$ TeV and, its holds true for both NH and IH.  

\begin{figure}[h]

	\begin{subfigure}[b]{0.45\textwidth}
		\centering
		\includegraphics[width=\linewidth]{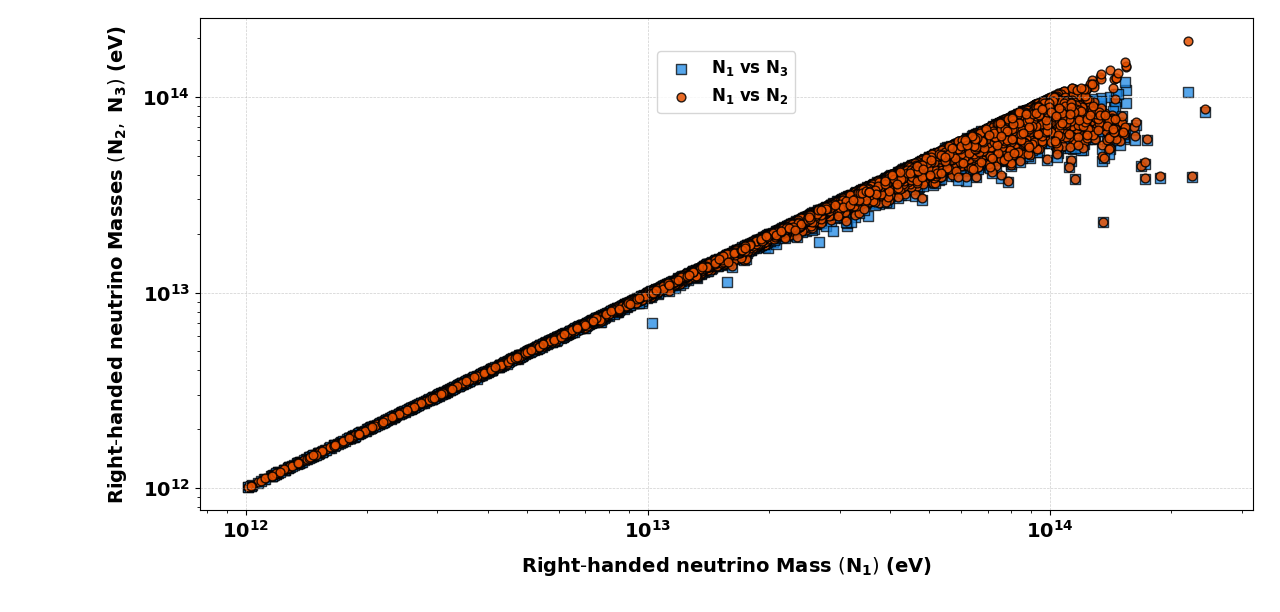}
		\caption{For NH}
		\label{f19}
	\end{subfigure}
	\hfill
	\begin{subfigure}[b]{0.45\textwidth}
		\centering
		\includegraphics[width=\linewidth]{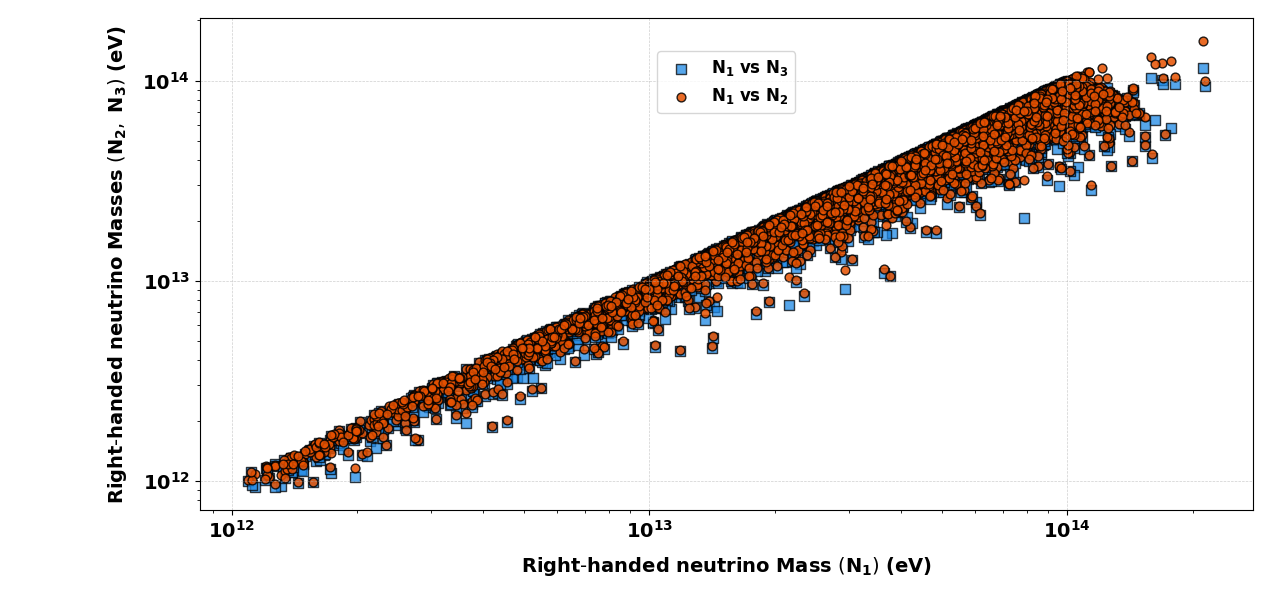}
		\caption{For IH}
		\label{f20}
	\end{subfigure}
	\caption{The two Figures show the parameter space for RH neutrino masses. }
\end{figure}
\begin{figure}[h]

	\begin{subfigure}[b]{0.45\textwidth}
		\centering
		\includegraphics[width=\linewidth]{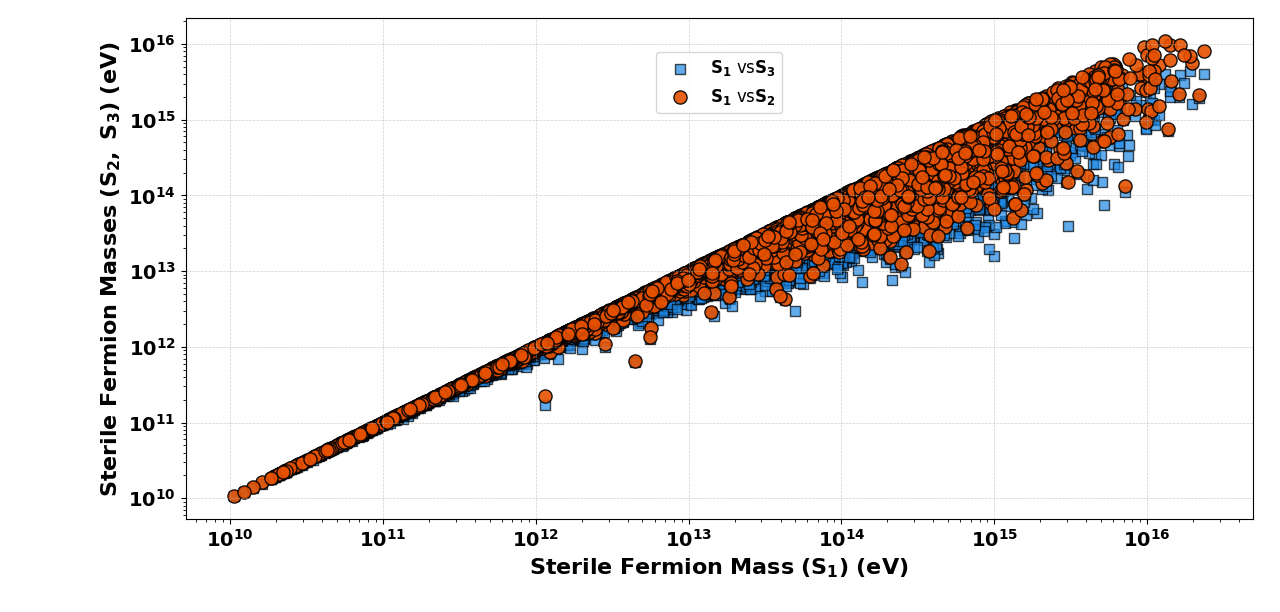}
		\caption{For NH }
		\label{f23}
	\end{subfigure} 
	\hfill
	\begin{subfigure}[b]{0.45\textwidth}
		\centering
		\includegraphics[width=\linewidth]{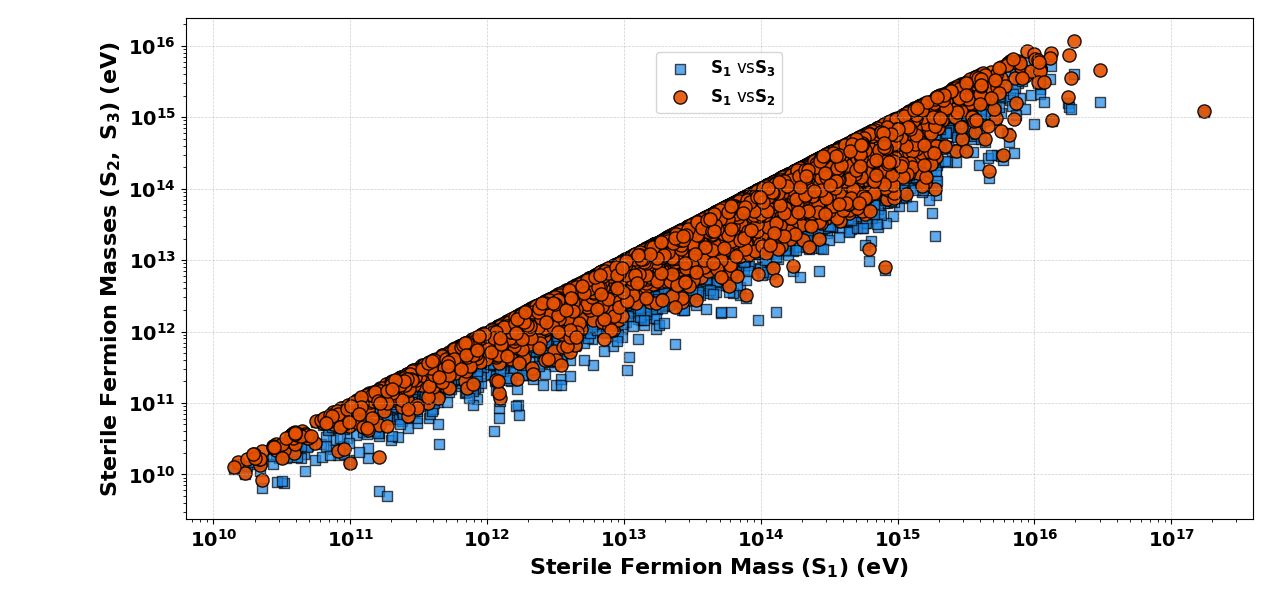}
		\caption{For IH}
		\label{f24}
	\end{subfigure} 
	\caption{The two Figures show the parameter space for sterile fermion masses. }
\end{figure}

\subsection{Effective neutrino mass and Half life of $0\nu\beta\beta$ decay} 
In Section VII, we have briefly discussed all the contributions to $0\nu\beta\beta$ decay associated with our model and in this section, we have discussed effective mass and its corresponding half-life and done a detailed numerical analysis of all the contributions. We have plotted graphs showing the variation of the effective mass of $0\nu\beta\beta$ decay and its corresponding half-life of ($0\nu\beta\beta$) decay with the lightest neutrino mass, considering both standard and non-standard contributions. In all the plots, which represent the variation of effective mass with the lightest neutrino mass, the presence of a horizontal line indicates the Planck constraint on the total sum of neutrino masses, while the two vertical lines represent the uncertainty on the effective mass due to nuclear mass matrix and effective mass should be less than this bound. Similarly, in all the figures, related to the half-life of $0\nu\beta\beta$ the horizontal line represents the Planck constraint, whereas the two vertical lines indicate the experimental bounds on the half-life for the isotope $Xe^{136}$. Different experimental bounds on the half-life of $0\nu\beta\beta$ for the isotopes Ge and Xe are given in Table \ref{TAB11}. As already mentioned within the left-right asymmetric framework, gauge coupling, $g_{l}$ and $g_{r}$ are not equal and their values depend upon the symmetry-breaking chain. Due to an unequal gauge coupling, we can see that an extra term $(\frac{g_{r}}{g_{l}})$ appears in the calculation of effective Majorana mass and half-life of $0\nu\beta\beta$ decay. We have considered the ratio $\frac{g_{r}}{g_{l}}$ equal to $0.6$ \cite{Pritimita:2016fgr} and the right-handed gauge boson mass $M_{W_{R}}$ of $3$ TeV. A detailed discussion about the effective Majorana mass and half-life of $0\nu\beta\beta$ is given in the Appendix \ref{AA} and \ref{AB} respectively. 
\begin{table}[h]
	\begin{tabular}{|c|c|c|c|c|}
		\hline
		Isotope &	$76_{Ge}$ & $136_{Xe}$ & 	$136_{Xe}$ & $136_{Xe}$ \\ \hline
	 $T^{0\nu}_{1/2}[years]$ & $>1.8 \times 10^{26}$ &$3.5 \times 10^{25}$ &$>1.9 \times 10^{25}$& $>2.3 \times 10^{26}$ \\ \hline
	 Experiment	 & GERDA \cite{Garfagnini:2024rvs}  & EXO \cite{EXO-200:2019rkq} & KamLAND-Zen \cite{KamLAND-Zen:2022tow} & EXO + KamLAND-Zen \cite{KamLAND-Zen:2012mmx} \\ \hline
	\end{tabular}
	\caption{Experimental bounds on half-life}
	\label{TAB11}
\end{table} 
\subsubsection{Effective mass and half-life due to $W_{L}-W_{L}$ current}
\begin{figure}[h]
	\begin{subfigure}[b]{0.45\textwidth}
		\centering
		\includegraphics[width=\linewidth]{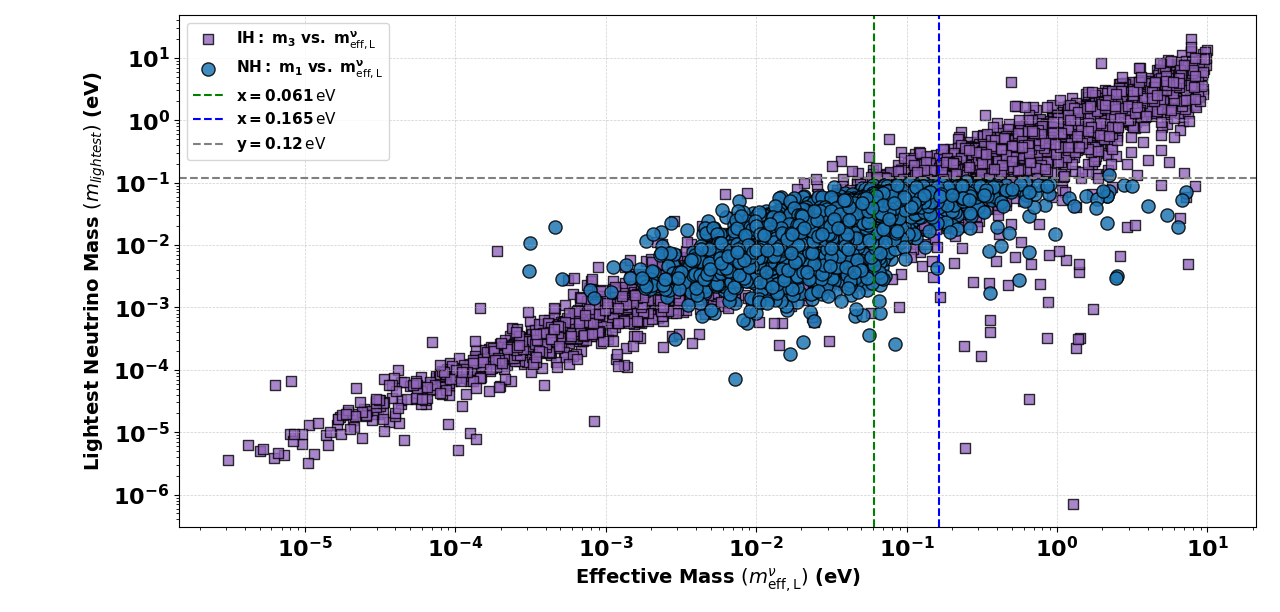}
		\caption{$m_{eff}$ due to light neutrino exchange}
		\label{f25}
	\end{subfigure} 
	\hfill
	\begin{subfigure}[b]{0.45\textwidth}
		\centering
		\includegraphics[width=\linewidth]{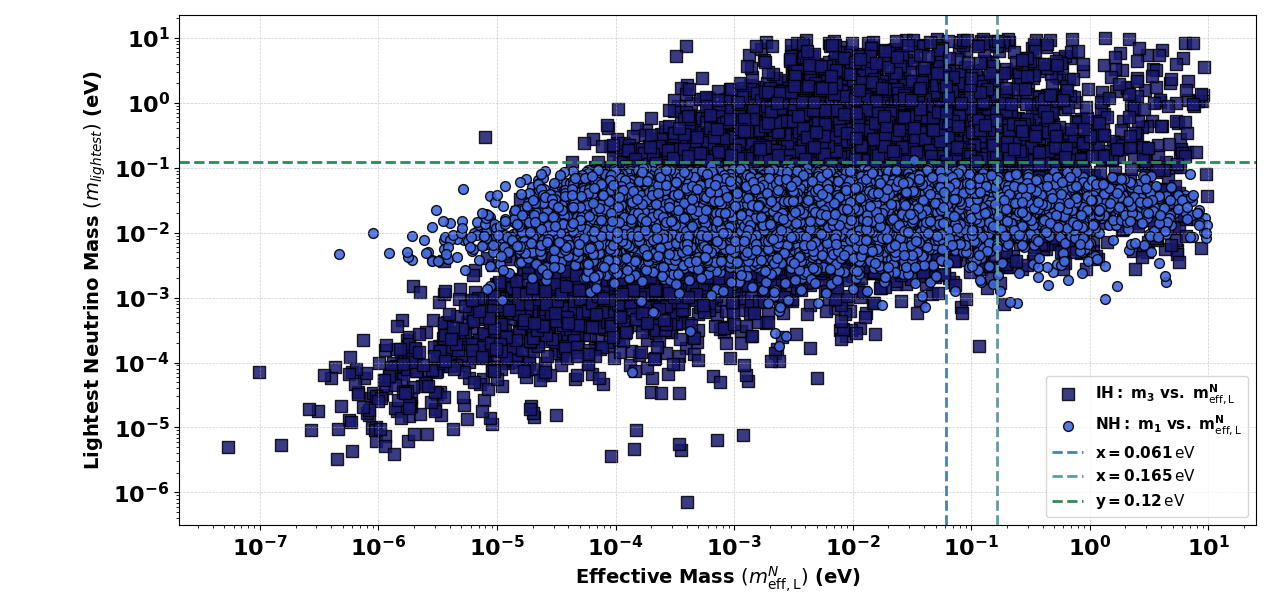}
		\caption{$m_{eff}$ due to RH neutrino exchange}
		\label{f26}
	\end{subfigure}
	
	\begin{subfigure}[b]{0.45\textwidth}
		\centering
		\includegraphics[width=\linewidth]{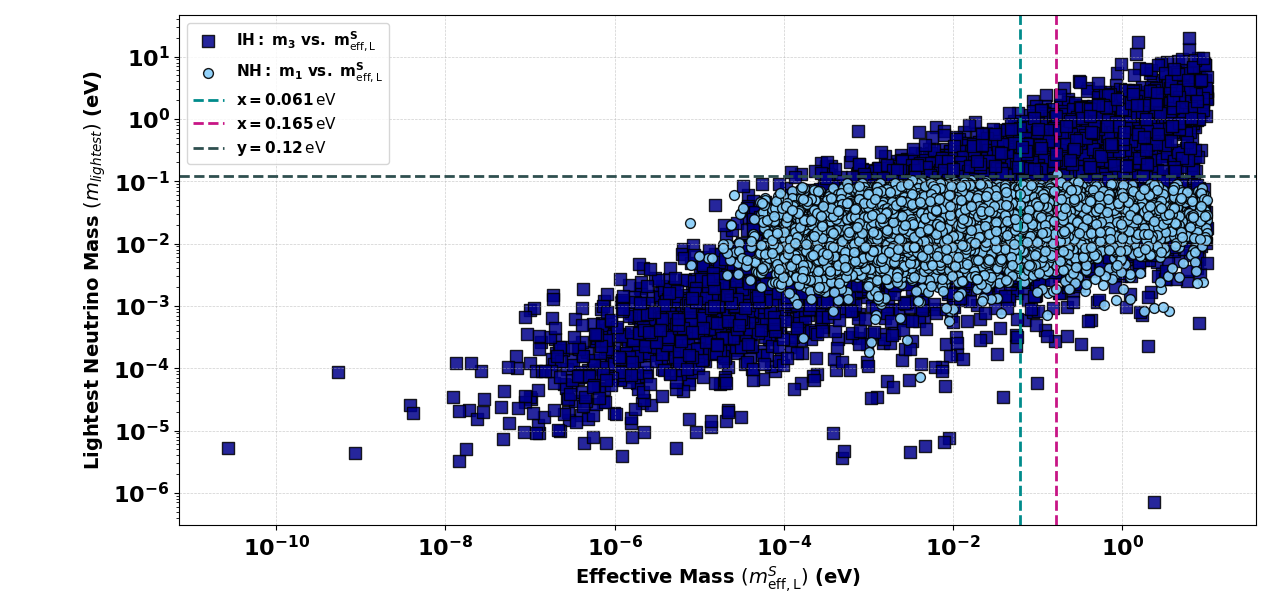}
		\caption{ $m_{eff}$ due to sterile fermion exchange}
		\label{f27}
	\end{subfigure} 

	\begin{subfigure}[b]{0.45\textwidth}
		\centering
		\includegraphics[width=\linewidth]{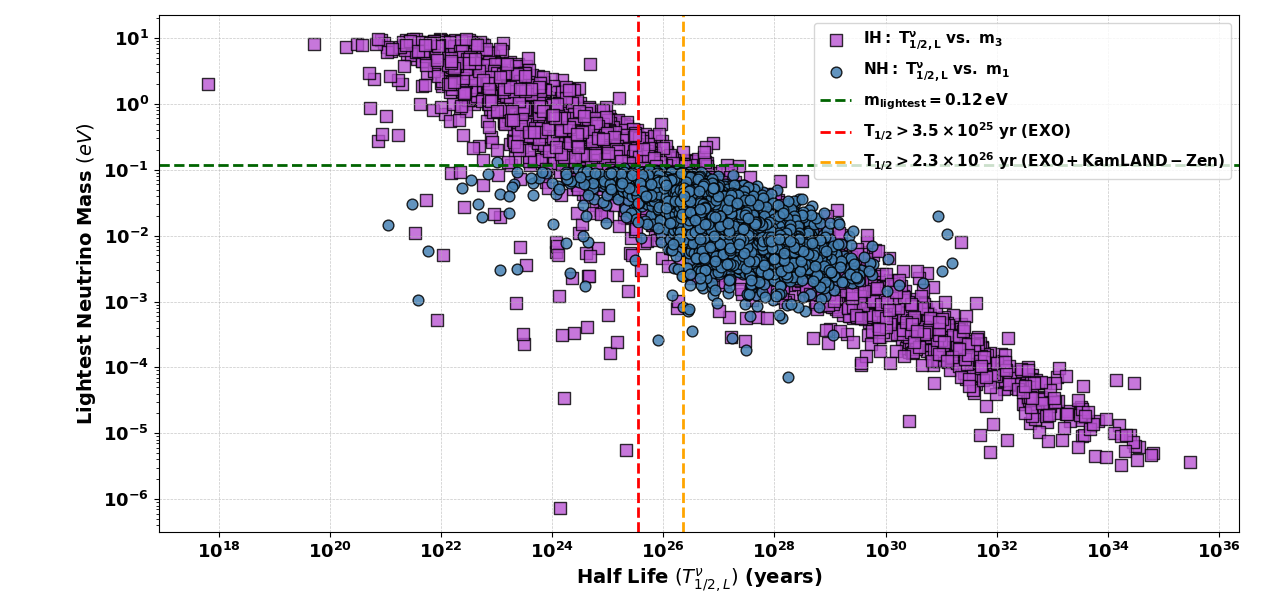}
		\caption{Half life due to standard contribution}
		\label{f28}
	\end{subfigure} 
	\hfill
	\begin{subfigure}[b]{0.45\textwidth}
		\centering
		\includegraphics[width=\linewidth]{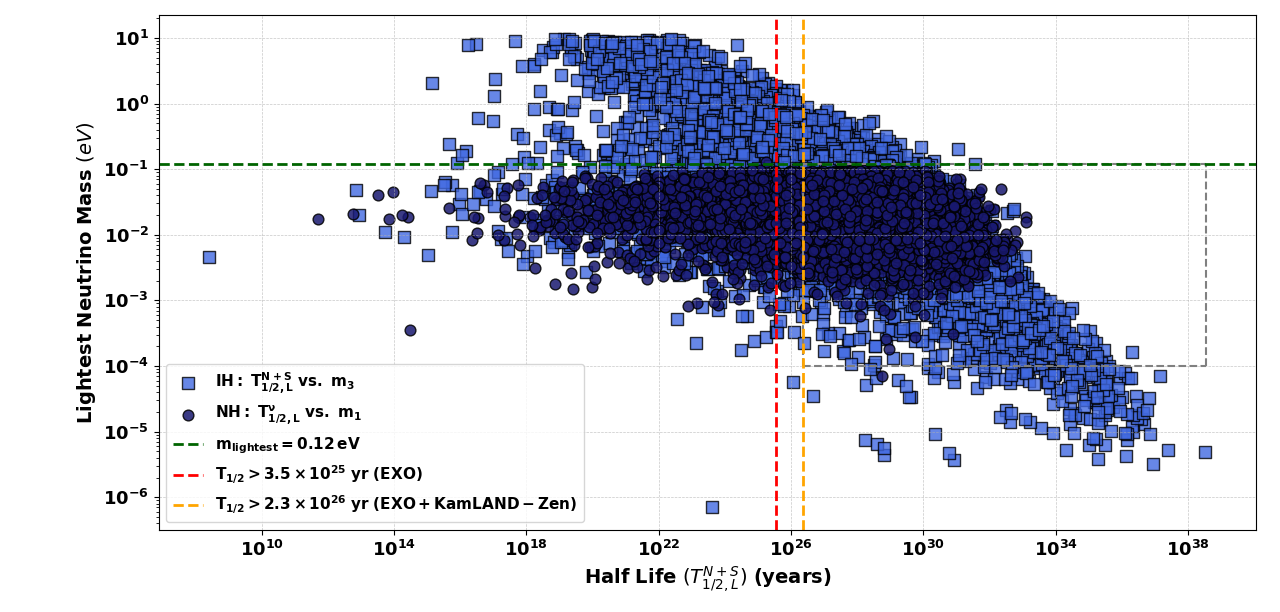}
		\caption{Half life due to non standard contribution }
		\label{f29}
	\end{subfigure}
	\caption{The top three Figures show the variation of effective mass with the lightest neutrino mass for NH and IH. The bottom two Figures show the variation of Haff-life of $0\nu\beta\beta$ decay with lightest neutrino mass for NH and IH }
	\label{fig5}
\end{figure}
The effective Majorana mass and half life due to exchange of light neutrino and heavy neutrino in case of $W_{L}-W_{L}$ current are given in the equation \eqref{qq3} and \eqref{qq4} respectively

\begin{equation} \label{qq3}
m^{\nu}_{ee,L} = \sum^{3}_{i=1}V^{\nu\nu^{2}}_{ei}m_{\nu i}, ~~~
m^{N}_{ee,L} =\sum^{3}_{i=1}V^{\nu N^{2}}_{ei} \frac{|p|^{2}}{M_{N_{i}}}, ~~
m^{S}_{ee,L}= \sum^{3}_{i=1} V^{\nu S^{2}}_{ei}\frac{|p|^{2}}{M_{S_{i}}}~~~. \\
\end{equation}
\begin{equation}\label{qq4}
 	[T^{0\nu}_{1/2}]^{-1} = \mathcal{K}_{0\nu} \Big[ |m^{\nu}_{ee,L}|^{2} + |m^{S}_{ee,L}+ m^{N}_{ee,L}|^{2} \Big]
 \end{equation}
We have calculated the effective Majorana mass and its corresponding half-life of $0\nu\beta\beta$ for $W_{L}-W_{L}$ mediation due to the exchange of light neutrino, RH neutrino, and sterile neutrino and plotted those calculated values against lightest neutrino mass. In Figures \ref{f25}, \ref{f26}, and \ref{f27}, we have observed that most of the calculated values of the effective mass due to light, heavy RH, and sterile neutrino exchange fall well below the experimental bound. In the case of NH, the model predicts the lower bound on the effective Majorana mass to be of the order of $10^{-3}$ to $10^{-4}$ when the exchange particle is a light or sterile neutrino, and approximately of the order of $10^{-6}$ when the exchange particle is a RH neutrino. However, in the case of IH, the lower bound predicted by the model due to the exchange of light neutrino, heavy RH neutrino, and sterile neutrino is approximately of the order of $10^{-5}$ to $10^{-8}$, which is smaller compared to the NH case. We have also calculated the half-life of $0\nu\beta\beta$ decay due to light neutrino exchange and due to the exchange of heavy neutrinos for NH and IH, which is shown in Figure \ref{f28} and \ref{f29} respectively.   
\subsubsection{Effective mass and half-life due to $W_{R}-W_{R}$ current}
The effective Majorana mass and half life due to exchange of light neutrino and heavy neutrino in case of purely right handed current are given in the equation \eqref{qq5} and \eqref{qq6} respectively
\begin{gather}\label{qq5}
	\begin{aligned}
		m^{\nu}_{ee,R}& =  \Bigg(\frac{M_{W_L}}{M_{W_R}}\Bigg)^{4}\Bigg(\frac{g_{r}}{g_{l}}\Bigg)^{4}\sum^{3}_{i=1}V^{N\nu^{2}}_{ei}m_{\nu i}\\ 
		m^{N}_{ee,R}& = \Bigg(\frac{M_{W_L}}{M_{W_R}}\Bigg)^{4}\Bigg(\frac{g_{r}}{g_{l}}\Bigg)^{4}\sum^{3}_{i=1}V^{NN^{2}}_{ei} \frac{|p|^{2}}{M_{N_{i}}}\\ 
		m^{S}_{ee,R}& =  \Bigg(\frac{M_{W_L}}{M_{W_R}}\Bigg)^{4}\Bigg(\frac{g_{r}}{g_{l}}\Bigg)^{4}\sum^{3}_{i=1} V^{N S^{2}}_{ei}\frac{|p|^{2}}{M_{S_{i}}}\\
	\end{aligned}
\end{gather}
\begin{equation}\label{qq6}
	[T^{0\nu}_{1/2}]^{-1} = \mathcal{K}_{0\nu} \Big[|m^{\nu}_{ee,R} + m^{S}_{ee,R} + m^{N}_{ee,R}|^{2}\Big]    
\end{equation}
\begin{figure}[h]

	\begin{subfigure}[b]{0.45\textwidth}
		\centering
		\includegraphics[width=\linewidth]{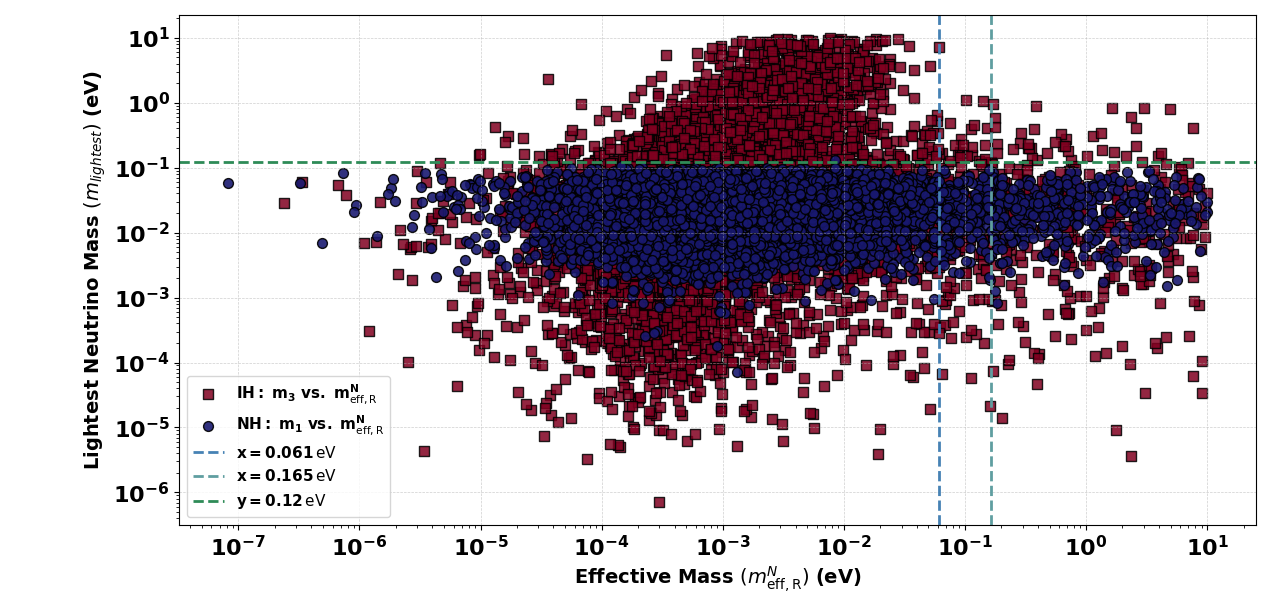}
		\caption{$m_{eff}$ due to the exchange of RH neutrino}
		\label{f30}
	\end{subfigure} 
	\hfill
	\begin{subfigure}[b]{0.45\textwidth}
		\centering
		\includegraphics[width=\linewidth]{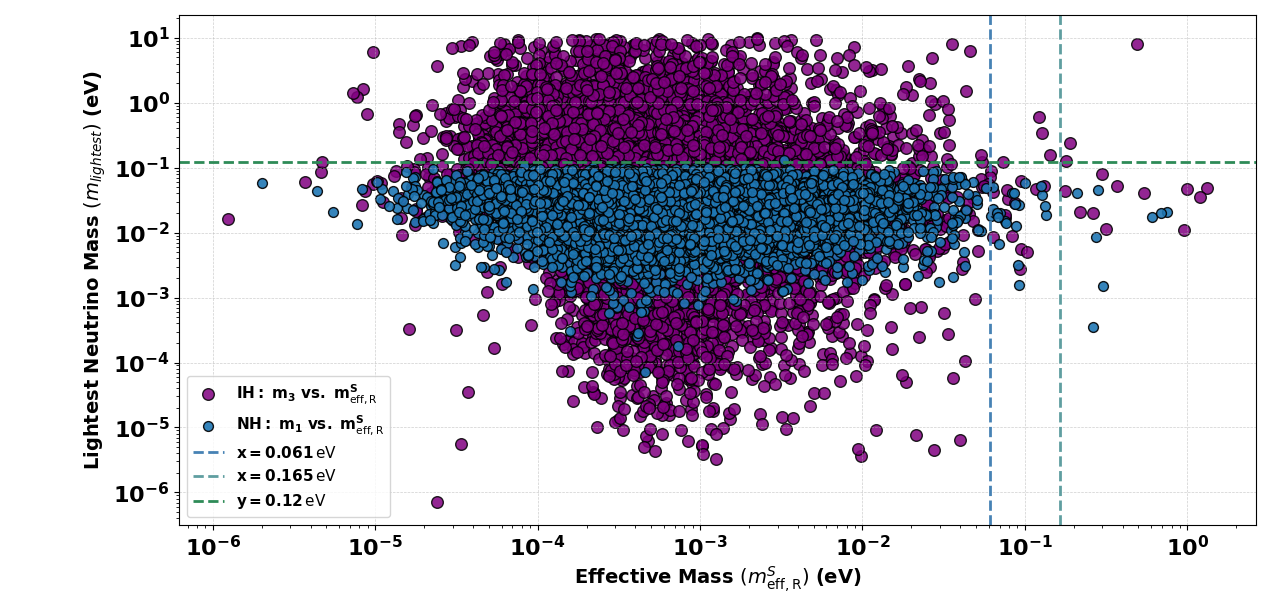}
		\caption{$m_{eff}$ due to the exchange of sterile fermion}
		\label{f31}
	\end{subfigure}
	
	\begin{subfigure}[b]{0.45\textwidth}
		\centering
		\includegraphics[width=\linewidth]{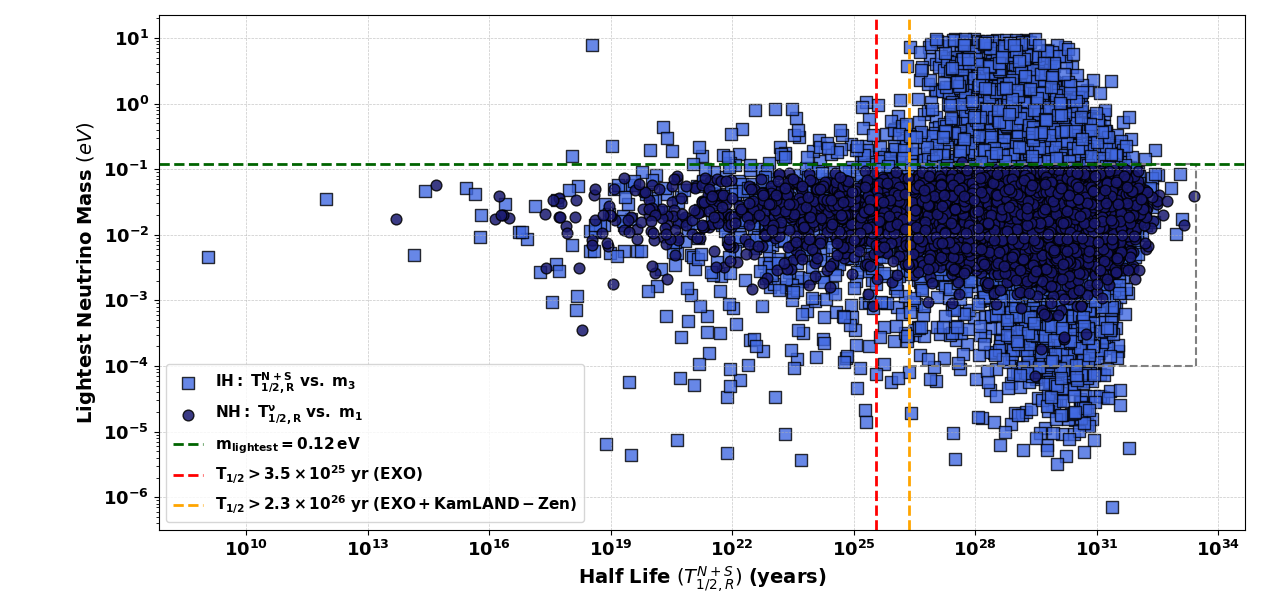}
		\caption{Half life due to non standard contribution}
		\label{f32}
	\end{subfigure} 
	\caption{The top two figures show the variation of effective mass with the lightest neutrino mass  for NH and IH. The bottom Figure show the variation of half-life with the lightest neutrino mass for NH and IH}
	\label{fig9}
\end{figure}

We have calculated the effective Majorana mass and corresponding half-life for the $( 0\nu\beta\beta)$ decay mediated by $W_{R}-W_{R}$ interactions, considering the exchange of light neutrinos, RH, and sterile neutrinos. In Figures \ref{f30} and \ref{f31}, we observed that the calculated effective masses due to RH and sterile neutrino exchanges consistently fall below the experimental bounds. Notably, the contribution from the exchange of light neutrinos via $W_{R}-W_{R}$ currents was found to be significantly smaller compared to other mechanisms, resulting in a very large corresponding half-life. Given this negligible contribution, we have omitted it from our total half-life calculations related to $W_{R}-W_{R}$ interactions. In Figure \ref{f32}, we have depicted the calculated half-lives for ($0\nu\beta\beta$) decay resulting from heavy neutrino exchanges, plotted against the lightest neutrino mass for both NH and IH scenarios.
\subsection{Calculation of non-unitary matrix and $J_{CP}$}
\begin{table}[h]
	\centering
	\renewcommand{\arraystretch}{1.5}
	\begin{tabular}{|c|c|c|c|c|}
		\hline
		\textbf{$\eta_{ij}$} & \textbf{NH Min} & \textbf{NH Max} & \textbf{IH Min} & \textbf{IH Max} \\ \hline
		$\eta_{ee}$         & $9.94 \times 10^{-8}$  & $2.3 \times 10^{-3}$  & $8.92 \times 10^{-11}$  & $2.4 \times 10^{-2}$  \\ \hline
		$\eta_{e\mu}$       & $3.9 \times 10^{-9}$   & $1.2 \times 10^{-3}$  & $3.2 \times 10^{-11}$   & $2.4 \times 10^{-2}$  \\ \hline
		$\eta_{e\tau}$      & $1.7 \times 10^{-9}$   & $1.2 \times 10^{-3}$  & $3.5 \times 10^{-11}$   & $2.1 \times 10^{-2}$  \\ \hline
		$\eta_{\mu\mu}$     & $8.8 \times 10^{-8}$   & $2.3 \times 10^{-3}$  & $1.5 \times 10^{-10}$   & $5.5 \times 10^{-2}$  \\ \hline
		$\eta_{\mu\tau}$    & $1.02 \times 10^{-9}$  & $2.2 \times 10^{-3}$  & $3.8 \times 10^{-11}$  & $2.9 \times 10^{-2}$  \\ \hline
		$\eta_{\tau\tau}$   & $8.8 \times 10^{-8}$   & $3.8 \times 10^{-3}$  & $1.7 \times 10^{-10}$   & $8.1 \times 10^{-2}$  \\ \hline
	\end{tabular}
	\caption{Minimum and maximum values of the elements of the non-unitary matrix $\eta_{ij}$ for NH and IH.}
	\label{TAB12}
\end{table}
\begin{table}[h]
	\centering
	\renewcommand{\arraystretch}{1.5}
	\begin{tabular}{|c|c|c|c|c|}
		\hline
		\textbf{$\Delta J^{ij}_{\alpha\beta}$} 
		& \textbf{NH (Min)} 
		& \textbf{NH (Max)} 
		& \textbf{IH (Min)} 
		& \textbf{IH (Max)} \\ \hline
		
		$\Delta J^{21}_{e\mu}$    
		& $-5.1\times 10^{-4}$ & $8.9\times 10^{-6}$ 
		& $-3.3 \times 10^{-3}$ & $1.5 \times 10^{-3}$ \\ \hline
		
		$\Delta J^{13}_{e\mu}$    
		& $-2.2\times 10^{-4}$ & $8.7\times 10^{-6}$ 
		& $-2.3 \times 10^{-3}$ & $6.4 \times 10^{-4}$ \\ \hline
		
		$\Delta J^{12}_{e\tau}$    
		& $-1.2\times 10^{-5}$ & $1.1\times 10^{-4}$ 
		& $-5.7 \times 10^{-4}$ & $1.6 \times 10^{-3}$ \\ \hline
		
		$\Delta J^{13}_{\mu e}$    
		& $-5.9\times 10^{-6}$ & $2.5\times 10^{-4}$ 
		& $-7.6 \times 10^{-4}$ & $2.3 \times 10^{-3}$ \\ \hline
		
		$\Delta J^{12}_{\mu\tau}$   
		& $-1.2\times 10^{-4}$ & $7.3\times 10^{-5}$ 
		& $-1.7 \times 10^{-3}$ & $1.8 \times 10^{-3}$ \\ \hline
		
		$\Delta J^{31}_{\tau e}$   
		& $-2.2\times 10^{-4}$ & $2.4\times 10^{-5}$
		& $-1.6 \times 10^{-3}$ & $1.6 \times 10^{-4}$ \\ \hline
		
	\end{tabular}
	\caption{Calculated minimum and maximum values of $\Delta J^{ij}_{\alpha \beta}$ for both NH and IH.}
	\label{TAB13}
\end{table}
Due to the presence of heavy RH and sterile neutrinos we obtain a non-unitary PMNS matrix and the term $\eta=\frac{1}{2}XX^{\dag}$ represents the deviation from unitary. Our calculations show that the deviations from unitarity, represented by the elements of the $\eta$ matrix, predominantly lie within the range of $10^{-3}$ to $10^{-9}$, which remain well below the current experimental bounds for the NH. For the IH, the calculated values span a broader range, from $10^{-2}$ to $10^{-11}$. Table \ref{TAB12} presents the computed lower and upper bounds for each element of the matrix.
Examining the CP violation terms resulting from non-unitarity, we find that for NH,  $J^{13}_{e\mu} $ is equal to $J^{32}_{e\mu}$  and their values range from $-2.1 \times 10^{-4}$ to $8.7\times 10^{-6}$. We have also get equal value for $J^{31}_{e\mu}$ and $J^{23}_{e\mu}$  and their values vary from $-8.7 \times 10^{-6}$ to $2.1 \times 10^{-4}$. For IH, we find that $J^{13}_{e\mu}$ values range from $-2.3\times 10^{-3}$ to $6.4\times 10^{-4}$ and $J^{31}_{e\mu}$ values range from $-6.4\times 10^{-4}$ to $2.3\times 10^{-3}$. Additionally, we observe that the absolute values of $J^{ij}_{e\tau}$ are equal. A Similar result is also observed for $J^{ij}_{\mu\tau}$, where $i \neq j$ and it can take values from $1$ to $3$. The Calculated CP violation term due to non-unitary effects are given in Table \ref{TAB13}.
\section{\textbf{Conclusion}} \label{s9}
We have developed a non-supersymmetric model using the concept of modular symmetry, where the usual requirement of holomorphicity is replaced by the Laplacian condition. The Yukawa couplings in this model include both holomorphic and non-holomorphic parts. We have studied the left-right asymmetric model, where we have used the intermediate gauge group $SU(2)_{R} \times U(1)_{R} \times U(1)_{B-L}$ to explore the neutrino masses and mixing by using the $\Gamma_{3}$ modular group, which is isomorphic to the $A_{4}$ discrete symmetric group. From the calculated Yukawa couplings obtained from the model for both NH and IH, we determined the real, imaginary, and absolute values of the modulus $\tau$ using the $q$-expansion of modular forms. We found that, in the case of NH, most of the values of real part of $\tau$ lies in the range $0.1$ to $0.3$, while the imaginary part falls within the upper half of the complex plane, ranging from $2.8\times 10^{-5}$ to $0.25$. For IH, the real part of $\tau$ lies between $0.1$ and $0.4$, and the imaginary part ranges from $-1.5\times10^{-5}$ to $0.2$. Additionally, the absolute value of $\tau$ is found to range from $0.117$ to $2.028$ in the case of NH, and from $0.109$ to $5.302$ in the case of IH. We have generated the light neutrino mass by using an extended inverse seesaw mechanism and for that purpose, we have introduced one sterile fermion per generation. The model predicts the sum of neutrino masses, with values well below experimental bounds for both NH and IH. The model predicts the lower bound on the lightest neutrino mass for NH to be approximately $10^{-4}$ eV, and around $10^{-5}$ eV for IH. In addition to neutrino masses, the model successfully predicts the mixing angles within the $3\sigma$ range. Furthermore, it predicts that the $\delta_{CP}$ phase is similar to one of the Majorana phases, denoted as $\alpha$ in the model. Additionally, we calculated effective mass and half-lives due to the mediation of $W_{L}-W_{L}$, and $W_{R}-W_{R}$ contribution. We have observed that the analytical formula for effective mass gets modified for $W_{R}-W_{R}$ mediation and depend on the ratio $\frac{g_{r}}{g_{l}}$. All the calculated values of the effective mass parameter and its corresponding half-lives due to standard and non-standard contributions are found to lie in the experimental allowed region for both the NH and IH cases. From our analysis, we find that in the case of the $W_L$–$W_L$ current, the model predicts the effective Majorana mass to lie within the range of $10^{-3}$~eV to $10^{-4}$~eV, and the half-life of the $0\nu\beta\beta$ decay to fall within the range of $10^{28}$ to $10^{30}$ years for the NH. In contrast, for the IH, the effective Majorana mass is predicted to lie within the range of $10^{-5}$~eV to $10^{-6}$~eV, with the corresponding half-life ranging from $10^{32}$ to $10^{34}$ years. Similarly, in the case of the $W_R$–$W_R$ current, the model predicts that the effective Majorana mass, arising from the exchange of heavy RH neutrinos and sterile neutrinos, lies within the range of $10^{-4}$~eV to $10^{-5}$~eV for both NH and IH. The corresponding half-life of the $0\nu\beta\beta$ decay is found to lie in the range of $10^{31}$ to $10^{32}$ years, which holds true for both hierarchies. Furthermore, we analyzed deviations from unitarity in the relevant matrix elements, finding them within the range of $10^{-3}$ to $10^{-9}$.\\
In conclusion, our non-supersymmetric left-right asymmetric model, based on the framework of automorphic forms, successfully predicts neutrino oscillation parameters and the effective mass and half-life of neutrinoless double beta decay.

\revappendix
\section{ Feynman diagram and Feynman amplitude for different contribution to $0\nu\beta\beta$ decay}\label{AA} 
\subsubsection{Contribution due to $W_{L}-W_{L}$ current}
\begin{figure}[h]
	\centering
	\includegraphics[width=\linewidth]{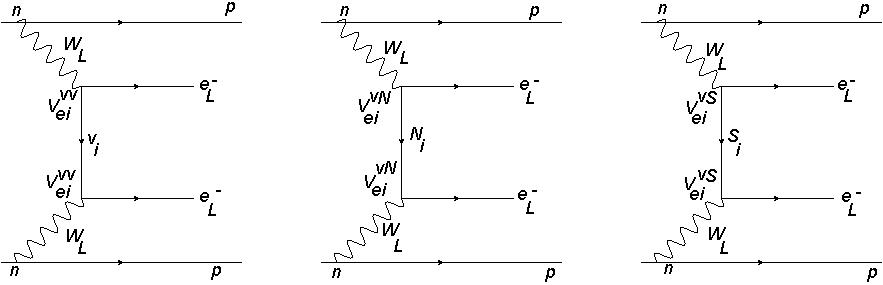}
	\caption{Feynman diagram for $W_{L}-W_{L}$ mediation}
	\label{Fig1}
\end{figure}

 The Feynman diagram associated with those contributions are given in Figure \ref{Fig1} and the associated Feynman amplitude and its corresponding lepton number violating (LNV) particle physics parameter are given in the equation \eqref{Q25}, and \eqref{Q26} respectively.
\begin{equation}\label{Q25}
	\mathcal{A}^{\nu}_{LL}  \propto G^{2}_{R} \sum_{i=1,2,3} \frac{V^{\nu\nu}_{ei}m_{\nu_i}}{p^{2}}, ~~~   \mathcal{A}^{N}_{LL}   \propto G^{2}_{F} \sum_{i=1,2,3}(-\frac{V^{\nu N^2}_{ei}}{M_{N_i}}), ~~~ \mathcal{A}^{S}_{LL} \propto G^{2}_{F} \sum_{i=1,2,3}(-\frac{V^{\nu S^2}_{ek}}{M_{S_k}})
\end{equation}
where  $G_{F}=1.2\times 10^{-5} GeV^{-2}$ is the Fermi coupling constant.
\begin{equation}\label{Q26}
	|\eta^{\nu}_{LL}| = \sum_{i=1,2,3} \frac{V^{\nu\nu^2}_{ei}m_{\nu_i}}{m_{e}}, ~~ |\eta^{N}_{LL}|=m_{p}\sum_{i=1,2,3}\frac{V^{\nu N^2}_{ei}}{M_{N_i}}, ~~ |\eta^{S}_{LL}|=m_{p}\sum_{i=1,2,3} \frac{V^{\nu S^2}_{ei}}{M_{S_i}}
\end{equation}
\subsubsection{Contribution due to $W_{R}-W_{R}$ current}
\begin{figure}[h]
	\centering
	\includegraphics[width=\linewidth]{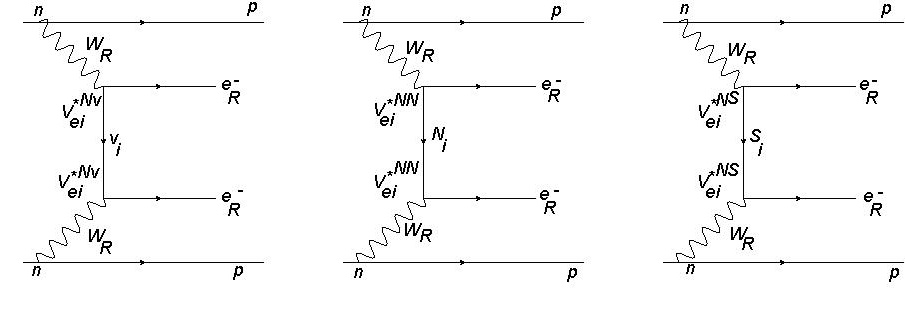}
	\caption{Feynman diagram for $W_{R}-W_{R}$}
	\label{Fig2}
\end{figure}

The Feynman diagram for those contributions are given in Figure \ref{Fig2}. Equations\eqref{Q27}  and \eqref{Q28} provide the amplitude of those diagrams and LNV parameters respectively.
\begin{gather}\label{Q27}
	\begin{aligned}
		\mathcal{A}_{RR}^{\nu} & \propto G^{2}_{F} \sum_{i=1,2,3} (\frac{M_{W_L}}{M_{W_R}})^{4} (\frac{g_{r}}{g_{l}})^{4} \frac{V_{ei}^{N\nu^2}m_{\nu_i}}{p^{2}} \\ 
		\mathcal{A}^{N}_{RR} & \propto G^{2}_{F} \sum_{J=1,2,3} (\frac{M_{W_L}}{M_{W_R}})^{4} (\frac{g_{r}}{g_{l}})^{4} (-\frac{V_{ej}^{N N^2}}{M_{N_j}}) \\
		\mathcal{A}^{S}_{RR} & \propto G^{2}_{F} \sum_{k=1,2,3} (\frac{M_{W_L}}{M_{W_R}})^{4} (\frac{g_{r}}{g_{l}})^{4} (-\frac{V_{ek}^{NS^2}}{M_{S_k}}) \\
	\end{aligned}
\end{gather}
\begin{gather}\label{Q28}
	\begin{aligned}
		|\eta|_{RR}^{\nu} & = \sum_{i=1,2,3} (\frac{M_{W_L}}{M_{W_R}})^{4} (\frac{g_{r}}{g_{l}})^{4} \frac{V_{ei}^{N\nu^2}m_{\nu_i}}{m_{e}} \\ 
		|\eta|^{N}_{RR} & = \sum_{i=1,2,3} m_{p} (\frac{M_{W_L}}{M_{W_R}})^{4} (\frac{g_{r}}{g_{l}})^{4} (\frac{V_{ei}^{N N^2}}{M_{N_i}}) \\
		|\eta|^{S}_{RR} & = \sum_{i=1,2,3} m_{p} (\frac{M_{W_L}}{M_{W_R}})^{4} (\frac{g_{r}}{g_{l}})^{4} (\frac{V_{ei}^{NS^2}}{M_{S_i}}) \\
	\end{aligned}
\end{gather}
\subsubsection{Contribution due to $W_{L}-W_{R}$ current}
\begin{figure}[h]
	\centering
	\includegraphics[width=\linewidth]{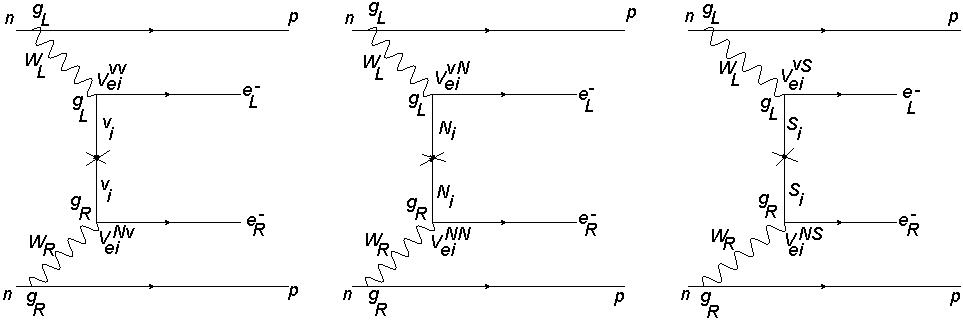}
	\caption{Feynman diagram for $\lambda$}
	\label{Fig3}
\end{figure}

\begin{figure}[h]
	\centering
	\includegraphics[width=\linewidth]{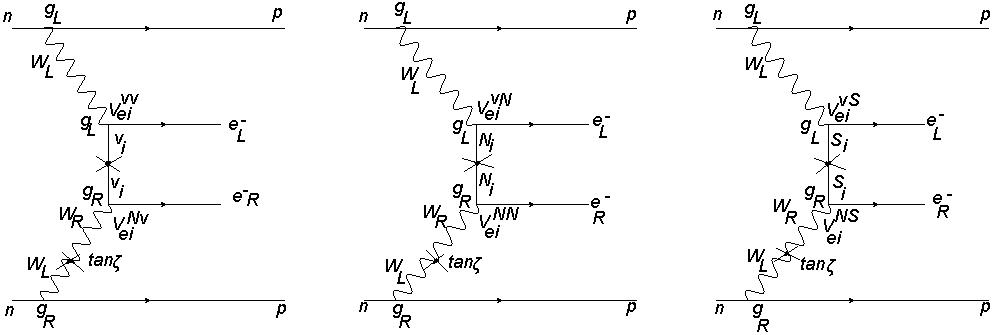}
	\caption{Feynman diagram for $\eta$}
	\label{Fig4}
\end{figure}

In case of $W_{L}-W_{R}$ mediation, we can have two type of mixed helicity Feynman diagram, which is known as the $\lambda$ and $\eta$ diagram. The Feynman diagram associated with those contribution is given in the Figure \ref{Fig3} and \ref{Fig4} respectively. 
\begin{itemize}
	\item[1] Feynman amplitudes for $\lambda$- mechanism 
	\begin{gather}\label{Q29}
		\begin{aligned}
			\mathcal{A}_{\lambda}^{\nu}& \propto G^{2}_{F} (\frac{M_{W_L}}{M_{W_R}})^{2} (\frac{g_{r}}{g_{l}})^{2} \sum_{i=1,2,3} V_{ei}^{\nu\nu}V_{ei}^{N\nu} \frac{1}{|p|} \\
			\mathcal{A}_{\lambda}^{N}& \propto G^{2}_{F} (\frac{M_{W_L}}{M_{W_R}})^{2} (\frac{g_{r}}{g_{l}})^{2} \sum_{j=1,2,3} V_{ej}^{\nu N}V_{ej}^{N N} \frac{|p|}{M^{2}_{N_j}} \\
			\mathcal{A}_{\lambda}^{N} & \propto G^{2}_{F} (\frac{M_{W_L}}{M_{W_R}})^{2} (\frac{g_{r}}{g_{l}})^{2} \sum_{k=1,2,3} V_{ek}^{\nu \nu}V_{ek}^{NS} \frac{|p|}{M^{2}_{S_k}} \\
		\end{aligned}
	\end{gather}    	
	
	\item[2] Feynman amplitudes for $\eta$-mechanism
	\begin{gather} \label{Q30}   
		\begin{aligned}
			\mathcal{A}_{\lambda}^{\nu} & \propto G^{2}_{F} (\frac{g_{r}}{g_{l}}) ten\zeta \sum_{i=1,2,3} V_{ei}^{\nu\nu}V^{N\nu}_{ei} \frac{1}{|p|} \\
			\mathcal{A}_{\lambda}^{N} & \propto G^{2}_{F} (\frac{g_{r}}{g_{l}}) ten\zeta \sum_{j=1,2,3} V_{ej}^{\nu N}V_{ej}^{N N} \frac{|p|}{M^{2}_{N_j}}  \\
			\mathcal{A}_{\lambda}^{S} & \propto G^{2}_{F} (\frac{g_{r}}{g_{l}}) ten\zeta \sum_{k=1,2,3} V_{ek}^{\nu S}V_{ek}^{N S} \frac{|p|}{M^{2}_{S_k}} \\
		\end{aligned}
	\end{gather}
\end{itemize}
\subsubsection{Contribution due to doubly charged Higgs boson}
\begin{figure}[h]
	\centering
	\includegraphics[width=\linewidth]{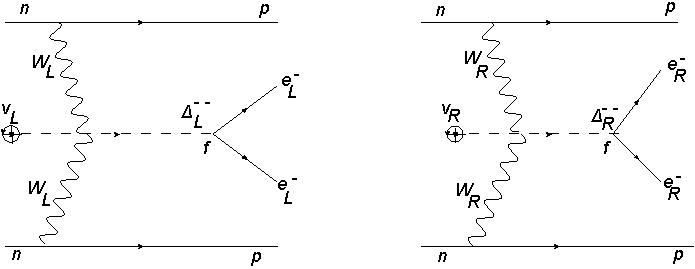}
	\caption{Feynman diagram for doubly charged Higgs boson}
	\label{Fig5}
\end{figure}

The Feynman amplitude and diagram due to doubly charged Higgs boson are given in equation \ref{qq31} and in Figure \ref{Fig5} respectively
	\begin{gather} \label{qq31}   
	\begin{aligned}
		\mathcal{A}^{\Delta_L}_{LL} & \propto G^{2}_{F} \sum_{i=1,2,3} \frac{1}{M^{2}_{\Delta_L}}V^{\nu\nu^2}_{ei}m_{\nu_i} \\
		\mathcal{A}^{\Delta_R}_{RR} & \propto G^{2}_{F} \sum_{i=1,2,3} \Big(\frac{M_{W_L}}{M_{W_R}}\Big)^{4} \Big(\frac{g_{r}}{g_{l}}\Big)^{4} \frac{1}{M_{\Delta_R}^{2}}V_{ei}^{NN^2}M_{N_i}
	\end{aligned}
\end{gather}
\section{Half-life of $0\nu\beta\beta$ decay}\label{AB}
The analytic expression for the inverse half life of neutrino less double beta decay considering all the contribution can be written in the following way \cite{Barry:2013xxa,Muto:1989cd}
\begin{equation}\label{Q31}
	[T_{1/2}^{\nu}]^{-1} = G^{0\nu}_{01} {|M^{0\nu}_{\nu}|^{2}|\eta^{\nu}_{LL}|^{2} + |M^{0\nu}_{N}|^{2}|\eta^{N,S}_{LL}|^{2}+ |M^{0\nu}_{\nu}|^{2}|\eta^{\nu}_{RR}|^{2}+|M^{0\nu}_{N}|^{2}|\eta^{N,S}_{RR}|+|M^{0\nu}_{\lambda}\eta_{\lambda}+M^{0\nu}_{\eta}\eta_{\eta}|^{2} }.
\end{equation}
Where $G^{0\nu}_{01}$ is the phase space factor and $M^{0\nu}_{i}(i=\nu, N, \lambda, \eta)$ is the nuclear matrix elements and $|\eta|$ is a dimensionless parameter. The experimental values of nuclear mass matrix for the isotope Ge and Xe is given in the Table \ref{TAB5}. Considering only the standard contribution due to the exchange of light Majorana neutrino we can write inverse half life formula in the following way \cite{Pritimita:2016fgr}
\begin{equation}\label{Q32}
	[T^{\nu}_{1/2}]^{-1}= G^{0\nu}_{01}|M^{0\nu}_{\nu}|^{2}|\eta^{\nu}_{LL}|^{2}
\end{equation} 
and by using equation \eqref{Q26}, one can modify equation \eqref{Q32} in terms of effective mass $|m^{\nu}_{ee,L}|$
\begin{equation}\label{q31}
	[T^{\nu}_{1/2}]^{-1}= G^{0\nu}_{01} |\frac{M^{0\nu}_{\nu}}{m_{e}}|^{2} |m^{\nu}_{ee,L}|^{2}
\end{equation}
where, $m_{e}$ is mass of the electron. One can take the term $\mathcal{K}_{0\nu}=G^{0\nu}_{01} |\frac{M^{0\nu}_{\nu}}{m_{e}}|^{2}$ as a normalizing factor for other contribution and can write equation \eqref{Q31} in  terms of the effective Majorana mass parameter in the following way

\begin{table}[h]
	\begin{tabular}{|c|c|c|c|c|c|}
		\hline
		Isotope & $G^{0\nu}_{01}$ & $M^{0\nu}_{\nu}$ & $M^{0\nu}_{N}$ & $M^{0\nu}_{\lambda}$& $M^{0\nu}_{\eta}$ \\ \hline
		$76_{Ge}$ & $5.77 \times 10^{-15}$ & $2.58-6.64$ & $233-412$ & $1.75-3.76$ & $235-637$\\ \hline
		$136_{Xe}$ & $3.56 \times 10^{-14}$ & $1.57-3.85$ & $164-172$ & $1.92-2.49$ & $370-419$ \\ \hline
	\end{tabular}
	\caption{Allowed ranges of phase space factors and nuclear matrix elements\cite{Senapati:2020alx}}
	\label{TAB5}
\end{table}
\begin{equation}\label{Q33}
	[T^{0\nu}_{1/2}]^{-1} = \mathcal{K}_{0\nu} \Big[ |m^{\nu}_{ee,L}|^{2} + |m^{S}_{ee,L}+ m^{N}_{ee,L}|^{2} + |m^{\nu}_{ee,R} + m^{S}_{ee,R} + m^{N}_{ee,R}|^{2} + |m^{\nu}_{ee,\lambda} + m^{S}_{ee,\lambda} + m^{N}_{ee,\lambda}|^{2} + |m^{\nu}_{ee,\eta} + m^{S}_{ee,\eta} + m^{N}_{ee,\eta}|^{2} \Big]           
\end{equation}  
\section{$A_{4}$ group}\label{AC}
$A_{4}$ is a  group which represents even permutation of four objects. It has a total of $12$ elements and have two generator, represented by $S$ and $T$. $A_{4}$ has $4$ conjugacy class which means that $A_{4}$ has four irreducible representation. Among those four irreducible representation, three are one dimensional $(1, 1^{\prime}, 1^{\prime\prime})$ and one is three dimensional. The product rules of two triplet ($\psi_{1}, \psi_{2}, \psi_{3}$) and ($\phi_{1}, \phi_{2}, \phi_{3}$) in the $T$ diagonal basis are given bellow
\begin{align*}
\begin{pmatrix}
\psi_{1}\\
\psi_{2}\\
\psi_{3}
\end{pmatrix} \times 
\begin{pmatrix}
\phi_{1}\\
\phi_{2}\\
\phi_{3}
\end{pmatrix}	= \begin{pmatrix}
                   \psi_{1}\phi_{1}+\psi_{2}\phi_{3}+\psi_{3}\phi_{2}
               \end{pmatrix} + \begin{pmatrix}
                 \psi_{3}\phi_{3}+\psi_{1}\phi_{2}+\psi_{2}\phi_{1}
               \end{pmatrix} + \begin{pmatrix}
                \psi_{2}\phi_{2}+\psi_{3}\phi_{1}+\psi_{1}\phi_{3}
               \end{pmatrix}  \\ 
               + \begin{pmatrix}
	2\psi_{1}\phi_{1}-\psi_{2}\phi_{3}-\psi_{3}\phi_{2} \\
	2\psi_{3}\phi_{3}-\psi_{1}\phi_{2}-\psi_{2}\phi_{1}	 \\
		2\psi_{2}\phi_{2}-\psi_{1}\phi_{3}-\psi_{3}\phi_{1}
	\end{pmatrix} +
	\begin{pmatrix}
	\psi_{2}\phi_{3}-\psi_{3}\phi_{2} \\
	\psi_{1}\phi_{2}-\psi_{2}\phi_{1} \\
	\psi_{3}\phi_{1}-\psi_{1}\phi_{3} 
	\end{pmatrix} ~~.
\end{align*}
The multiplication rules are given bellow
\begin{equation*}
3\times 3 = 1 + 1^{\prime} + 1^{\prime\prime} + 3 + 3.
\end{equation*}
\begin{equation*}
1^{\prime} \times 1^{\prime} =1^{\prime\prime}, ~~ 1^{\prime} \times 1^{\prime\prime} =1, ~~ 1^{\prime\prime} \times 1^{\prime\prime} =1^{\prime}.
\end{equation*}
\bibliographystyle{unsrt}
\bibliography{cite}
	
\end{document}